\documentclass[letterpaper,journal]{IEEEtran}
\usepackage{amsmath,amsfonts}
\usepackage{algorithmic}
\usepackage{algorithm}
\usepackage{array}
\usepackage[caption=false,font=footnotesize]{subfig}
\usepackage{textcomp}
\usepackage{stfloats}
\usepackage{url}
\usepackage{verbatim}
\usepackage{graphicx}
\usepackage{cite}
\usepackage{amssymb}
\usepackage{acro}
\usepackage{balance}
\usepackage{xcolor}
\usepackage{booktabs}
\usepackage{tabularx}
\usepackage{array}
\usepackage{makecell}

\usepackage[table]{xcolor}
\makeatletter
\renewcommand{\fnum@figure}{\textbf{\figurename~\thefigure}}
\makeatother

\DeclareAcronym{ue}{
  short = UE,
  long  = user equipment,
}
\DeclareAcronym{mimo}{
  short = MIMO,
  long  = multiple-input multiple-output,
}
\DeclareAcronym{thz}{
  short = THz,
  long  = Terahertz,
}
\DeclareAcronym{ap}{
  short = AP,
  long  = access point,
}
\DeclareAcronym{upa}{
  short = UPA,
  long  = uniform planar array,
}
\DeclareAcronym{snr}{
  short = SNR,
  long  = signal-to-noise ratio,
}
\DeclareAcronym{rmse}{
  short = RMSE,
  long  = root-mean-squared error,
}
\DeclareAcronym{se}{
  short = SE,
  long  = spectral efficiency,
}
\DeclareAcronym{bd}{
  short = BD,
  long  = beam depth,
}
\DeclareAcronym{cca}{
  short = CA,
  long  = circular array,
}
\DeclareAcronym{csa}{
  short = SA,
  long  = square array,
}

\begin{document}

\title{Limitation of Bessel Beams in Near Field Communications:\\A Depth of Field Perspective}

\author{Aditya Jolly,~\IEEEmembership{Graduate Student Member,~IEEE},
Vitaly Petrov,~\IEEEmembership{Member,~IEEE},\\
Gábor Fodor,~\IEEEmembership{Fellow,~IEEE}, and Emil Björnson,~\IEEEmembership{Fellow,~IEEE}        

\thanks{A. Jolly, V. Petrov, G. Fodor, and E. Björnson are with KTH Royal Institute of Technology, Sweden. Email: \{ajolly, vitalyp, gaborf, emilbjo\}@kth.se. A. Jolly and G. Fodor are also with Ericsson Research, Sweden. }
}

\maketitle

\begin{abstract}
Bessel beams have recently been explored as a candidate complement to focused beams in radiative near-field communications due to their non-diffracting and self-healing properties. However, recent work has shown that the practically useful range of finite-aperture Bessel beams can occupy only a fraction of the radiative near-field region, establishing the existence of a finite propagation limit. In this paper, we establish such a limitation analytically from a depth-of-field perspective for Bessel beams generated by finite circular and square apertures. To quantify the Bessel beam depth, we first derive analytical expressions for the normalized array gain and use them to establish absolute and peak-normalized $3$-dB depth-of-field criteria. These criteria reveal critical cone-angle thresholds that partition the Bessel-beam configuration space into (i)~undefined, (ii)~finite, and (iii)~infinite depth-of-field regimes. Our analysis further reveals that the maximum usable non-diffracting range under the proposed depth-of-field criterion is limited to $\approx22\%$ and $\approx19\%$ of the Fraunhofer distance for circular and square apertures, respectively. Numerical results confirm the accuracy of the derived limits for discrete planar arrays, providing practical engineering guidelines for the design of finite-aperture Bessel beams in future near-field communications.
\end{abstract}

\begin{IEEEkeywords}
Radiative Near Field, MIMO, Array Gain, Bessel Beams, Beam Depth.
\end{IEEEkeywords}

\section{Introduction}
\IEEEPARstart{T}{he} ever-growing demand for higher data rates and network capacity has driven wireless systems toward higher frequencies, where larger bandwidths can be exploited~\cite{ITU_M2160}. However, increasing the signal frequency also introduces greater propagation losses, motivating the use of electrically large antenna arrays to maintain a sufficient link budget via highly directional transmissions. As wavelength decreases, while the physical size of the antenna arrays stays in the cm-scale, the so-called ``radiative'' near-field region can extend over tens or even hundreds of meters~\cite{tutorial1}. This effect is particularly profound for millimeter wave (mmWave) ($30$ -- $300~\mathrm{GHz}$), and especially at \ac{thz} (\(0.3\)--\(3~\mathrm{THz}\)) frequencies, where many future wireless links are expected to operate within the near-field regime~\cite{proc_ieee_jornet}.

Unlike in far-field scenarios, where electromagnetic propagation can be accurately described by planar waves, propagation in the near field becomes different~\cite{review_THz}. Consequently, using conventional far-field beamforming based on plane-wave assumptions in the near field may result in significant performance degradation~\cite{Perf}. This has recently motivated growing interest in near-field specific beamforming~\cite{BeamManipulation}. A diverse family of techniques has been proposed with distinct properties, including beamfocusing~\cite{review_THz}, Bessel beams~\cite{BB_Gen}, Mathieu beams~\cite{W_and_M}, Airy beams~\cite{airy}, Hermite--Gaussian beams~\cite{H_Gauss}, vortex beams~\cite{Vortex_beam_alignment}, among others. However, the finite aperture of practical antenna arrays can limit the region over which these beams retain their intended characteristic properties. Therefore, understanding these limits is important for assessing the practical use of these beams in the radiative near field.

These limitations have been partially explored. Among others, in~\cite{Primer_Emil}, a depth-of-field limitation was investigated for \emph{beamfocusing}, revealing that the finite-depth focusing is achievable only for receiver locations within approximately \emph{10\% of the Fraunhofer distance} from a \ac{upa}. Beyond this limit, the focused beam progressively loses its finite-depth behavior and effectively converges toward conventional far-field beamforming. These findings have informed the development of multi-user beamforming algorithms~\cite{Beamopt}, range-angle beam training~\cite{bTRAINbf}, codebook design~\cite{CB_NFBF}, and near-field channel estimation and localization techniques~\cite{E_chan_est_loc}.

This $\approx10\%$ limit for beamfocusing naturally raises the question of whether similar finite-depth limits exist for other beam types in the near field. Among the various near-field beams mentioned above, \emph{Bessel beams}~\cite{BB_Gen} have gained significant attention due to their defining propagation characteristics, namely the ability to maintain their beam profile over the propagation direction (\emph{non-diffracting})~\cite{BB_Old} and reconstruct after partial obstruction (\emph{self-healing})~\cite{wavefront_eng}. These properties make Bessel beams particularly attractive for providing robustness to user-location uncertainty~\cite{BB_6G,spec_manage,ULA_NF_Beams_limit} due to their non-diffracting property, and for blockage-resilient communications~\cite{spec_manage,ULA_NF_Beams_limit,opt_mrt_BB} due to their self-healing capability. This has spurred extensive research on Bessel beams, spanning beam synthesis, configuration design, experimental demonstrations, and localization applications, highlighting their potential to improve spectral efficiency in ultra-wideband \ac{thz} systems~\cite{BB_Gen,jollyVTC,Nature_Vitaly_bessel_beams,nearLoc}. 

The growing maturity of Bessel beam-based communication systems makes understanding their propagation limits increasingly important. This motivates the following research question: \emph{Do finite-aperture Bessel beams also exhibit (i)~the same, (ii)~similar, or (iii)~principally different propagation limit as beamfocusing?}

Importantly, recent work in~\cite{Dist_wave} provides evidence that such a limit may exist. The authors empirically measure the so-called ``Effective Near-Field Range (ENFR)'' of a Bessel beam, defined as a set of communication ranges where the Bessel beam shows at least a $3$~dB average advantage in received power over conventional far-field beamforming. However, as the work in~\cite{Dist_wave} primarily aims to empirically confirm the existence of such a limit, the authors do not provide any closed-form expressions for how to derive the limit for different Bessel beam configurations. Further, as the utilized ENFR metric averages the received power over the entire non-diffracting region, it does not directly reveal where exactly along the propagation direction the Bessel beam actually begins to lose its finite-depth behavior and deteriorate or how this transition depends on the beam configuration. This motivates the following questions explored in the present work: \emph{What determines the limiting distance of a finite-aperture Bessel beam, how does this limit constrain its feasible configuration space, and what is the maximum non-diffracting range over which a finite depth of field can be maintained?}

To answer these questions, this paper builds upon the normalized array-gain criterion consistent with the ones originally developed for beamfocusing. Importantly, unlike ENFR, the normalized array gain retains the spatial evolution of the beam along the propagation direction, allowing us to characterize when its finite-depth behavior is lost. We use this evolution to identify the feasible range of Bessel-beam configurations, ultimately determining the maximum usable non-diffracting range of a Bessel beam for near-field communications.

The main contributions of this paper are thus as follows: 
\begin{itemize}

\item We derive analytical expressions for the normalized array gain of a phase-based Bessel beam generated by circular and square arrays. Building upon these expressions, we introduce absolute and peak-normalized $3$-dB beam-depths and establish the conditions governing their existence and finiteness.

\item Using the derived existence and finiteness conditions, we obtain \emph{critical analytical thresholds that partition the Bessel beam configuration space into regimes with: (i)~undefined, (ii)~finite, and (iii)~infinite depth-of-field} for the \ac{cca} and \ac{csa} cases. We hence provide a complete characterization of the transition between different Bessel beam-depth regimes.

\item Building upon the derived analytical boundary conditions, we establish the quantitative limit on the maximum usable non-diffracting range of a Bessel beam in near-field communications. We show that this limit: (i)~is determined by the antenna aperture, geometry, and the signal frequency; and (ii)~corresponds to only a fraction of the Fraunhofer distance ($\approx20\%$). \emph{Moreover, through numerical simulations we demonstrate the accuracy of the derived analytical limits for discrete antenna arrays}.

\end{itemize}

The rest of the paper is organized as follows. Section~\ref{sec:syst_model} presents the scenario and the near-field channel model. Section~\ref{sec:NAG_BB} presents the normalized array gain for the \ac{cca} and \ac{csa} cases. Section~\ref{sec:dF_BB} defines the depth-of-field criterion and formulates the critical conditions. Section~\ref{sec:critical_angle_characterization} utilizes the critical conditions to derive the corresponding Bessel beam configuration ranges and establish the maximum usable non-diffracting range. Section~\ref{sec:numerical_results} utilizes numerical simulations and shows the impact of the established analytical limit on practical Bessel beam based near-field links. Finally, Section~\ref{sec:conclusion} concludes the paper. The key notations are summarized in Table~\ref{tab:notations}.

\begin{figure}[t!]
\centering

\subfloat[Circular array (\ac{cca}) with the radius $R$]{%
\includegraphics[width=0.96\columnwidth]{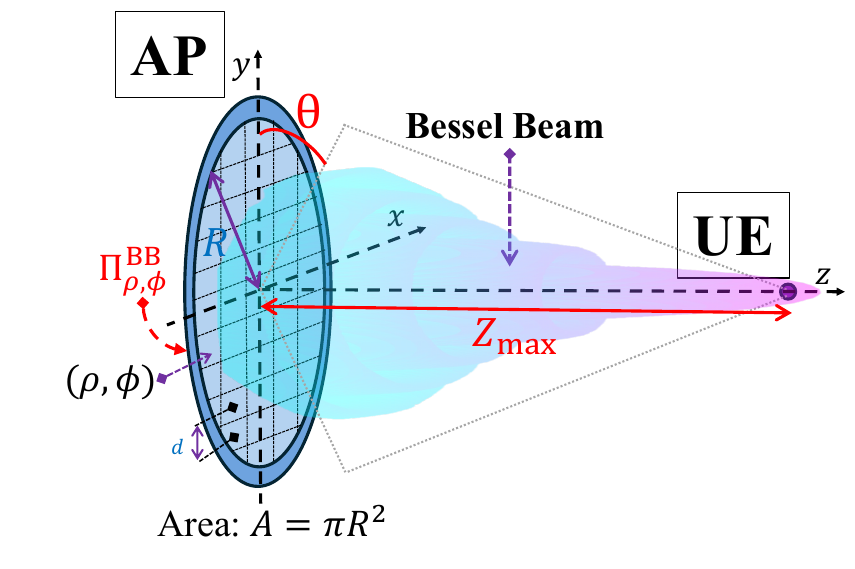}
\label{fig:Syst_Model_circ}}
\hfill

\subfloat[Square array (\ac{csa}) with the side length $D_{\mathrm{AP}}$]{%
\includegraphics[width=0.96\columnwidth]{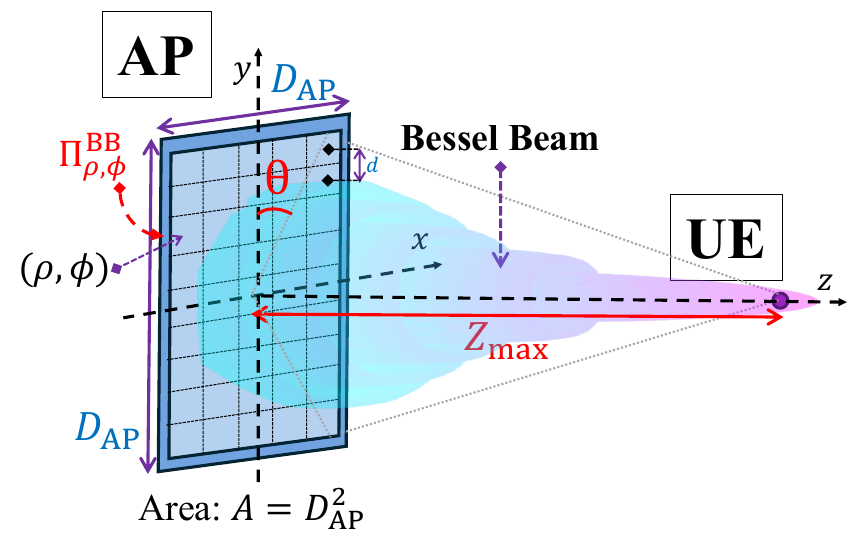}
\label{fig:Syst_Model_sq}}

\caption{Downlink near-field \ac{ap}--\ac{ue} communication system utilizing a Bessel phase profile ($\Pi_{\rho,\phi}^{\mathrm{BB}}$) with cone angle $\theta$ and non-diffracting range ($Z_{\max}$).}
\label{fig:Syst_Model}
\vspace{-14pt}
\end{figure}

\begin{table}[!]
\caption{List of Notations}
\label{tab:notations}
\centering
\scriptsize
\renewcommand{\arraystretch}{1.08}
\setlength{\tabcolsep}{3pt}
\begin{tabularx}{\columnwidth}{
>{\centering\arraybackslash}p{0.26\columnwidth}
!{\vrule width 0.4pt}
>{\raggedright\arraybackslash}X}
\toprule
\textbf{Notation} & \textbf{Description} \\
\midrule

\multicolumn{2}{c}{\textbf{\textit{System and Bessel-beam notation}}} \\
\midrule
$\lambda$ & Carrier wavelength \\
$z$ & \ac{ue} location measured from the array center \\
$\theta$ & Cone angle of the phase-based Bessel beam \\
$\Pi^{\rm BB}$ & Bessel-beam phase profile applied at the array \\
$Z_{\max}$ & Non-diffracting range, $Z_{\max}=R_{\rm eff}/\tan(\theta)$ \\
$h(\rho,\varphi, z)$ & Channel coefficient between the \ac{ue} and the point $(\rho,\varphi)$ on the \ac{ap} \\
$P_{\rm rx}(\theta,z)$ & Total received power at the \ac{ue} \\

\midrule
\multicolumn{2}{c}{\textbf{\textit{Array geometry and metrics}}} \\
\midrule
$A$ & Area; $A=\pi R^2$ for the \ac{cca} and $A=D_{\rm AP}^{2}$ for the \ac{csa} \\
$\rho_{\max}$ & Maximum radial distance from the origin to the array boundary \\
$R_{\rm eff}$ & Effective aperture radius \\
$G_{\rm norm}(\theta,z)$ & Normalized array gain \\
$\mathrm{BD}^{\rm abs}_{\rm 3dB}(\theta)$ & Absolute 3-dB beam depth \\
$\mathrm{BD}^{\rm rel}_{\rm 3dB}(\theta)$ & Peak-normalized 3-dB beam depth \\
\midrule
\multicolumn{2}{c}{\textbf{\textit{Circular array}}} \\
\midrule
$R$ & Radius of the \ac{cca} \\
$d^{c}_{\rm RA}$, $d^{c}_{\rm FA}$ & Reactive near-field boundary and Fraunhofer distance \\
$Z^{c}_{\max}$ & Non-diffracting range of a Bessel beam \\
$\mu_c$ & Normalized conic-phase parameter, $\mu_c=4R\sin(\theta)/\lambda$ \\
$\tau_c$ & Normalized distance, $\tau_c=z/d^{c}_{\rm FA}$ \\
$G^{c}_{\rm norm}$ & \ac{cca} normalized array gain \\
$\theta^c_{\rm crit,A1}, \theta^c_{\rm crit, A2}$ & Critical cone angle associated with $\mathrm{BD}^{\rm abs}_{\rm 3dB}(\theta)$\\
$\theta^c_{\rm crit, B}$ & Critical cone angle associated with $\mathrm{BD}^{\rm rel}_{\rm 3dB}(\theta)$\\
\midrule
\multicolumn{2}{c}{\textbf{\textit{Square array}}} \\
\midrule
$D_{\rm AP}$ & Side length of the \ac{csa} \\
$d^{s}_{\rm RA}$, $d^{s}_{\rm FA}$ & Reactive near-field boundary and Fraunhofer distance \\
$Z^{s}_{\max}$ & Non-diffracting range of a Bessel beam \\
$\mu_s$ & Normalized conic-phase parameter, $\mu_s=2D_{\rm AP}\sin(\theta)/\lambda$ \\
$\tau_s$ & Normalized distance, $\tau_s=z/d^{s}_{\rm FA}$ \\
$G^{s}_{\rm norm}$ & \ac{csa} normalized array gain \\
$\theta^s_{\rm crit,A1}, \theta^s_{\rm crit, A2}$ & Critical cone angle associated with $\mathrm{BD}^{\rm abs}_{\rm 3dB}(\theta)$\\
$\theta^s_{\rm crit, B}$ & Critical cone angle associated with $\mathrm{BD}^{\rm rel}_{\rm 3dB}(\theta)$\\
\bottomrule
\end{tabularx}
\end{table}

\section{System Model}
\label{sec:syst_model}

We consider a downlink communication scenario in which an \ac{ap} equipped with a planar array with inter-element spacing $d$ serves a \ac{ue} (point receiver) located in the radiative near-field of the transmitting array using a phase-based Bessel profile. We assume $d\geq\lambda/2$, and mutual coupling is ignored. We consider two array geometries: (i) \ac{cca} with radius $R$ and area $A=\pi R^2$, containing $N_c\approx\pi R^2/d^2$ elements arranged in a square grid (see Fig.~\ref{fig:Syst_Model_circ}), and (ii) \ac{csa} with side length $D_{\rm AP}$ and area $A=D_{\rm AP}^2$, containing $N_s\approx D_{\rm AP}^2/d^2$ elements (see Fig.~\ref{fig:Syst_Model_sq}).  For analytical tractability, the AP is represented using an ideal continuous spatial approximation of the array response. This assumption is justified due to the large number of antenna elements in electrically large arrays ($D \gg \lambda$), and is commonly used for analytical characterization~\cite{Primer_Emil}. The accuracy of this approximation is quantified in Section~\ref{subsec:discrete_array_valid}.

The \acp{ap} transmit with total power $P_{\rm tx}$ distributed equally across all antenna elements. Without loss of generality, both \ac{ap} geometries are considered to be parallel to the $xy$-plane and centered at the origin. The \ac{ue} is located on the \ac{ap} boresight and displaced only along the $z$-axis, enabling the analysis of the depth-of-field of a phase-based Bessel beam, as illustrated in Fig.~\ref{fig:Syst_Model}. Although the analysis is presented from the downlink perspective, the same formulation applies to the reciprocal uplink case by electromagnetic reciprocity~\cite{Reseprocity}.

\subsection{Near-field channel model}
\label{sec:Chanel_Model}

We adopt a line-of-sight (LoS), non-uniform spherical-wave near-field channel model~\cite{NF_primer_review}. We operate in the radiative near-field region of the array such that $d_{\rm RA}<z\leq d_{\rm FA}$. Here, $d_{\rm RA}$ and $d_{\rm FA}$ denote the standard
reactive and radiative near-field boundaries of the array, respectively~\cite{dfa} (see Section~\ref{sec:NAG_BB} for array-specific formulas).

Let $(\rho=\sqrt{x^2+y^2},\phi=\tan^{-1}(y/x),0)$, denote the point $(x,y,0)$ on the \ac{ap} \emph{in cylindrical coordinates}. The continuous spatial representation of the channel coefficient between the point $(\rho,\varphi)$ on the \ac{ap} and \ac{ue} located at $(0,0,z)$ is then
\begin{equation}
h(\rho,\phi,z)
=
\frac{\lambda}{4\pi r(\rho,z)}
e^{-j\frac{2\pi}{\lambda}r(\rho,z)},
\label{eq:channel_continuous}
\end{equation}
where $\lambda$ is the wavelength and $r(\rho,z)=\sqrt{\rho^2+z^2}$ denotes the distance between the point $({\rho,\phi})$ on the \ac{ap} and the \ac{ue}.

\subsection{Bessel beam generation}
\label{sec:BB_generation}
To generate a Bessel beam using a phase-based technique, a spatially varying phase profile $\Pi^{\rm BB}(\rho,\phi)$ is applied across the \ac{ap} aperture~\cite{BB_6G}:
\begin{equation}
\Pi^{\rm BB}(\rho,\phi)
=
\frac{2\pi}{\lambda}\rho\sin(\theta)
\label{eq:phase_diff_ele}
\end{equation}
Without loss of generality, the phase profile is referenced to the center of the array, i.e., the point $\rho=0$ is assigned zero phase. This phase-based technique applies a conic phase profile with cone angle $\theta$ at the transmitting array to generate a Bessel beam.
Its geometrical non-diffracting range is given by $Z_{\rm max}=R_{\rm eff}/\tan(\theta)$. Here, $R_{\rm eff}$ is the effective radius of
the \ac{ap} array. For \ac{cca}, $R_{\rm eff}=R$, while for \ac{csa}, $R_{\rm eff}=D_{\rm AP}/\sqrt{2}$, equivalent to the radius of the circumscribed circle~\cite{jollyVTC}.

\subsection{Received power}
\label{sec:Rec_pow}

The total received power observed at the \ac{ue} located at distance $z$ is obtained by coherently summing the contributions from all antenna elements on the \ac{ap}, which corresponds to an integral under the continuous array approximation. Then, with uniform power allocation across the array aperture ($A$), the received power $P_{\rm rx}$ at $z$ with Bessel phase profile (cone angle $\theta$) at the \ac{ap} is given by
\begin{equation}
P_{\rm rx}(\theta,z)
=
\frac{P_{\rm tx}}{A^{2}}
\left|
\int_{\mathcal{A}}
h(\rho,\phi,z)
e^{j\Pi^{\rm BB}(\rho,\phi)} \rho d\rho d\phi
\right|^2,
\label{eq:PRX_discrete}
\end{equation}
where $\mathcal{A}=\{(\rho,\phi,0): 0\leq\rho\leq\rho_{\max}(\phi),0\leq\phi\leq2\pi\}$ denotes the array aperture, with $\rho_{\max}(\phi)$ representing the radial distance from the origin to the aperture boundary.
\section{Normalized array gain of a phase-based Bessel beam}
\label{sec:NAG_BB}
In~\cite{Primer_Emil}, the normalized array gain is introduced to quantify the loss in coherent array gain in the radiative near field relative to the corresponding plane-wave array gain, given by
\begin{equation}
G_{\mathrm{norm}}(\theta,z)
=
\frac{
\left|
\int_{\mathcal{A}}
h(\rho,\phi,z)
e^{j\Pi^{\rm BB}(\rho,\phi)}
\rho\,d\rho\,d\phi
\right|^2
}{
A\int_{\mathcal{A}}
|h(\rho,\phi,z)|^2
\rho\,d\rho\,d\phi
}.
\label{eq:normalized_array_gain_general}
\end{equation}

Accordingly, $G_{\mathrm{norm}}=1$ corresponds to perfect phase alignment and maximum coherent array gain, while $G_{\mathrm{norm}}<1$ captures the loss due to phase mismatch across the aperture. Although~\cite{Primer_Emil} introduces this for an uplink scenario, the same metric applies for downlink to characterize the coherent field generated by the transmit array at a given observation point. 

The metric was originally developed for beamfocusing, for which $G_{\mathrm{norm}}=1$ can be attained at the focal point. However, both beamfocusing and Bessel beams rely on controlling the distribution of energy along the propagation direction. This makes $G_{\mathrm{norm}}$ a natural metric for quantifying the degradation of the coherent gain of Bessel beams with distance. Its evolution along the propagation direction is therefore used here to characterize the practical depth-of-field of finite-aperture Bessel beams. In the following subsections, we derive $G_{\mathrm{norm}}$ for the \ac{cca} and \ac{csa} cases.

\subsection{Normalized array gain of a circular array}
\label{subsec:CCA_gnorm}
Using the definition in~\eqref{eq:normalized_array_gain_general}, we first derive closed-form expression for the normalized array gain for a \ac{cca}. The exact gain can be calculated using numerical integration of the exact expression of $G_{\rm norm}$~\eqref{eq:normalized_array_gain_general}, where $\rho_{\max} = R$.
An analytical expression can be obtained using the Fresnel approximation~\cite{fresnel} $\sqrt{\rho^2+z^2}\approx z+\frac{\rho^2}{2z}$, which gives an error lower than $3.5\times 10^{-3}$ if $z \geq 2.5R$ \cite{Primer_Emil}, reducing the axial normalized gain to
\begin{equation}
    G^{c}_{\mathrm{norm}}(\theta,z)
    =
    \frac{4}{R^4}
    \left|
    \int_0^R
    \rho
    e^{-j\frac{2\pi}{\lambda}\left(\frac{\rho^2}{2z} - \rho\sin(\theta)\right)}
    d\rho
    \right|^2.
    \label{eq:gnorm_z_theta}
\end{equation}
\begin{figure*}[t]
\setcounter{equation}{9}
\begin{equation}
\begin{aligned}
G^{c}_{\mathrm{norm}}(\mu_c,\tau_c)
&=
(8\tau_c)^2
\Bigg|
\frac{j}{\pi}
\Bigg(
e^{
-j\frac{\pi}{2}
\left(
\frac{1}{2\sqrt{\tau_c}}
-
\mu_c\sqrt{\tau_c}
\right)^2}
-
e^{
-j\frac{\pi}{2}\mu_c^2\tau_c}
\Bigg)
-
\mu_c\sqrt{\tau_c}
\left(
\Delta C_{\rm c}(\mu_c,\tau_c)
-
j\Delta S_{\rm c}(\mu_c,\tau_c)
\right)
\Bigg|^2
\end{aligned}
\label{eq:gnorm_closed_form}
\end{equation}
\setcounter{equation}{13}
\begin{equation}
\small
\begin{aligned}
G^{s}_{\mathrm{norm}}(\mu_s,\tau_s)
&=
(16\tau_s)^2
\Bigg|
\int_{0}^{\pi/4}
\frac{j}{\pi}
\Bigg(
e^{
-j\frac{\pi}{2}
\left(
\frac{1}{2\sqrt{2\tau_s}\cos(\phi)}
-
\mu_s\sqrt{2\tau_s}
\right)^2}
-
e^{
-j\frac{\pi}{2}
\mu_s^2(2\tau_s)}
\Bigg)
-
\mu_s\sqrt{2\tau_s}
\left(
\Delta C_{\mathrm{s}}(\phi,\mu_s,\tau_s)
-
j\Delta S_{\mathrm{s}}(\phi,\mu_s,\tau_s)
\right)
d\phi
\Bigg|^2
\end{aligned}
\label{eq:square_gnorm_semiclosed_form}
\end{equation}
\vspace{-1em}
\hrulefill
\end{figure*}
Now, to simplify \eqref{eq:gnorm_z_theta}, we define the radial integral $I$ such that $G^{c}_{\mathrm{norm}}(\theta,z) = 4 |I|^2/R^4$, and 
\setcounter{equation}{5}
\begin{equation}
    I
    =
    \int_0^R
    \rho
    e^{-j\alpha \rho^2+j\beta\rho}d\rho,
    \label{eq:app_integral_alpha_beta}
\end{equation}
where $\alpha=\pi/(\lambda z)$, and $\beta=2\pi\sin(\theta)/\lambda$. The phase term can be rewritten by completing the square. Then,
we use the substitution $u=\rho-z\sin(\theta)$.
The corresponding limits are $u_1=-z\sin(\theta)$, and $u_2=R-z\sin(\theta)$.
Using $\rho=u+z\sin(\theta)=u+\beta/(2\alpha)$, the integral becomes
\begin{equation}
\small
    I
    =
    e^{j\beta^2/(4\alpha)}
    \left(
    \int_{u_1}^{u_2}
    u e^{-j\alpha u^2}du
    +
    \frac{\beta}{2\alpha}
    \int_{u_1}^{u_2}
    e^{-j\alpha u^2}du
    \right).
    \label{eq:app_integral_split}
\end{equation}

The phase factor $e^{j\beta^2/(4\alpha)}$ has unit magnitude and is
independent of $\phi$, and therefore does not affect the final normalized
gain. The first integral is elementary, therefore,
\begin{equation}
    \int_{u_1}^{u_2}
    u e^{-j\alpha u^2}du
    =
    \frac{j}{2\alpha}
    \left(
    e^{-j\alpha u_2^2}
    -
    e^{-j\alpha u_1^2}
    \right).
    \label{eq:app_first_integral_limits}
\end{equation}

For the second integral, define $t=\sqrt{2\alpha/\pi} u$.
Then $\alpha u^2=\pi t^2/2$ and $du=\sqrt{\pi/2\alpha}\ dt$,
and hence
\begin{equation}
\small
\begin{aligned}
    \int_{u_1}^{u_2} e^{-j\alpha u^2}du
    &=
    \sqrt{\frac{\pi}{2\alpha}}
    \left(
    C(t_2)-C(t_1)
    -
    j\left(S(t_2)-S(t_1)\right)
    \right),
\end{aligned}
    \label{eq:app_second_integral}
\end{equation}
where $C(\cdot)$ and $S(\cdot)$ are the Fresnel integrals, defined as $C(x)=\int_0^x \cos\left(\pi t^2/2\right)dt$, and $S(x)=\int_0^x \sin\left(\pi t^2/2\right)dt$. We introduce the normalized variables $\mu_c=4R\sin(\theta)/\lambda$, and $\tau_c=z/d^c_{\mathrm{FA}}$, where $d^c_{FA} = 8R^2/\lambda$ is the Fraunhofer distance of the \ac{cca}, and $\tau_c$ measures the observation distance relative to the Fraunhofer distance and $\mu_c$ measures the strength of the conic phase across the aperture.

The normalized limits can be written as $t_1=-\mu_c\sqrt{\tau_c}$, and $t_2=\frac{1}{2\sqrt{\tau_c}}-\mu_c\sqrt{\tau_c}$. Using the odd symmetry of the Fresnel integrals, the Fresnel differences can be
written as $C(t_2)-C(t_1)=-\Delta C_{c}(\mu_c,\tau_c)$, and $S(t_2)-S(t_1)=-\Delta S_{c}(\mu_c,\tau_c)$. 
First, substituting~\eqref{eq:app_first_integral_limits} and~\eqref{eq:app_second_integral} into \eqref{eq:app_integral_split} yields the radial integral $I$. Substituting~\eqref{eq:app_integral_split} into \eqref{eq:gnorm_z_theta} and expressing the results in terms of the normalized variables gives the closed-form expression at the top of the page in~\eqref{eq:gnorm_closed_form}.
\subsection{Normalized array gain of a square array}
\label{subsec:CSA_gnorm}
For a \ac{csa}, the exact normalized array gain can similarly be evaluated through numerical integration of~\eqref{eq:normalized_array_gain_general}, where
\setcounter{equation}{10}
\begin{equation}
\rho_{\max}
=
\frac{D_{\rm AP}}
{2\max\left\{|\cos(\phi)|,|\sin(\phi)|\right\}}.    
\end{equation}
An analytical expression can be obtained by applying the same Fresnel approximation~\cite{fresnel}, which remains accurate across the array with errors below $3\times 10^{-3}$ when $z \geq 1.2 D_{\rm AP}$~\cite{Primer_Emil}, and exploiting the eightfold symmetry of the \ac{csa}. The normalized array gain reduces to 
\begin{equation}
\small
    G^{s}_{\mathrm{norm}}(\theta,z)
    =
    \frac{64}{D_{\rm AP}^4}
    \left|
    \int_{0}^{\frac{\pi}{4}}
    \int_{0}^{\frac{D_{\rm AP}}{2\cos(\phi)}}
    \rho
    e^{-j\frac{2\pi}{\lambda}\left(\frac{\rho^2}{2z}- \rho\sin(\theta)\right)}
    d\rho d\phi
    \right|^2.
    \label{eq:square_fresnel_gain_symmetric}
\end{equation} 
\noindent Now, to simplify \eqref{eq:square_fresnel_gain_symmetric}, we define the radial integral ($I(\phi)$) such that $G^{s}_{\mathrm{norm}}(\theta,z) = 64|\int^{\pi/4}_{0}I(\phi)d\phi|^2/D_{\rm AP}^4 $, given by
\begin{equation}
    I(\phi)
    =
    \int_0^{q(\phi)}
    \rho
    e^{-j\alpha \rho^2+j\beta\rho}d\rho,
    \label{eq:app_square_integral_alpha_beta}
\end{equation}
where $q(\phi) = D_{\rm AP}/(2\cos(\phi))$ for $0\leq \phi\leq\pi/4$. Applying a similar technique as the \ac{cca}, the phase term can be written by completing the square. 
Then, the integral admits the same form as the \ac{cca} case~\eqref{eq:app_integral_split}, with the only difference being the upper integration limit $u_2=q(\phi)-z\sin(\theta)$. We introduce similar normalized variables $\mu_s=2D_{\rm AP}\sin(\theta)/\lambda$, and $\tau_s=z/d^s_{\mathrm{FA}}$ where $d^s_{FA}=4D_{AP}^2/\lambda$ is the Fraunhofer distance of the square array, and $\tau_s$ and $\mu_s$ have similar interpretations as the \ac{cca} case. The normalized limits can then be written as $t_1=-\mu_s\sqrt{2\tau_s}$, and $t_2=\frac{1}{2\sqrt{2\tau_s}\cos(\phi)}-\mu_s\sqrt{2\tau_s}$. Using the odd symmetry of the Fresnel integrals, the Fresnel differences can be written as $C(t_2)-C(t_1)=-\Delta C_{s}(\phi,\mu_s,\tau_s)$, and $S(t_2)-S(t_1)=-\Delta S_{s}(\phi,\mu_s,\tau_s)$.

Now substituting~\eqref{eq:app_square_integral_alpha_beta} into~\eqref{eq:square_fresnel_gain_symmetric} and expressing the result in terms of the normalized variables gives the expression shown at the top of the page in~\eqref{eq:square_gnorm_semiclosed_form}. The key difference from the \ac{cca} case is that, for the \ac{csa}, the radial upper limit depends on the azimuth angle as $\rho_{\max}=D_{\rm AP}/(2\cos(\phi))$ over $0\leq\phi\leq\pi/4$. Consequently, the radial integral has a closed-form primitive, but the \ac{csa} gain retains an angular integral over $\phi$ needing numerical evaluation.

\subsection{Spatial evolution of the normalized array gain}
\begin{figure*}[!t]
\centering

\subfloat[$\theta=10.63^{\circ}$, finite depth-of-field]{%
\includegraphics[width=0.66\columnwidth]
{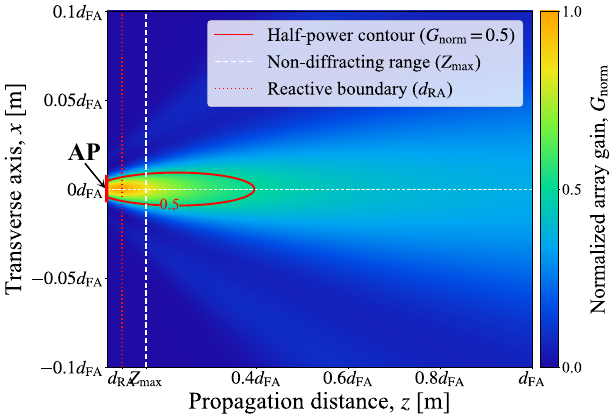}
\label{fig:bessel_beam_theta_2deg}}
\hfill
\subfloat[$\theta=8.63^{\circ}$, finite depth-of-field]{%
\includegraphics[width=0.66\columnwidth]
{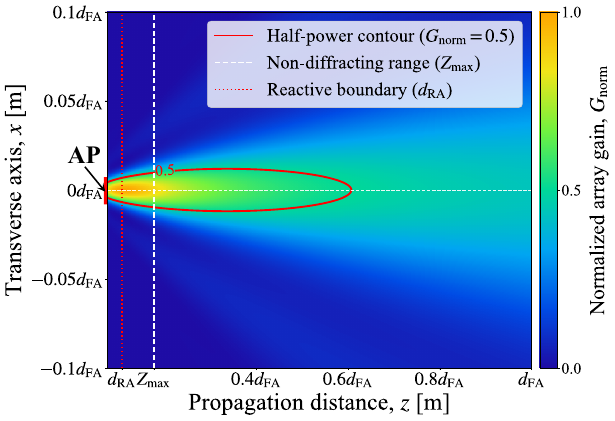}
\label{fig:bessel_beam_theta_1deg}}
\hfill
\subfloat[$\theta=7.63^{\circ}$, infinite depth-of-field]{%
\includegraphics[width=0.66\columnwidth]
{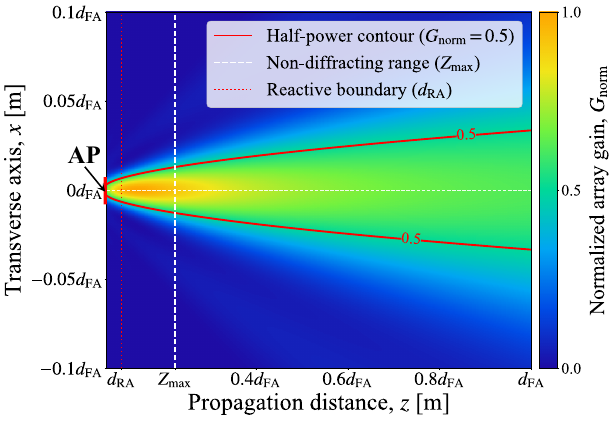}
\label{fig:bessel_beam_theta_01deg}}

\caption{Normalized array gain for a \ac{cca} with $R=10$~cm at $f=125$~GHz, illustrating the spreading behavior of finite-aperture Bessel beams for decreasing cone angle. Smaller cone angles produce a broader beam and a longer high-gain propagation region.}
\label{fig:bessel_beam_spreading_behavior}
\vspace{-12pt}
\end{figure*}

The derived expressions~\eqref{eq:gnorm_closed_form} and~\eqref{eq:square_gnorm_semiclosed_form} enable direct evaluation of the normalized array gain throughout the radiative near field. Fig.~\ref{fig:bessel_beam_spreading_behavior} illustrates the resulting spatial evolution of $G_{\rm norm}$. As the cone angle decreases, the high-gain region extends over a longer propagation distance while simultaneously broadening in the transverse direction. The figure also shows that contours of constant $0.5$ normalized array gain naturally show the effective high-gain propagation region. In particular, the $G_{\rm norm}=0.5$ ($3$~dB) contour (indicated in red) marks  the points at which the coherent array gain has decreased by $3$~dB relative to the plane-wave maximum.

These observations show that the conventional non-diffracting range $Z_{\max}$, while providing a geometric approximation of where the finite-aperture Bessel beam ceases to exist, does not capture how effectively the beam remains spatially confined and therefore does not directly characterize its communication performance. Instead, the normalized array gain reveals a gradual degradation in coherent gain as the cone angle decreases and the beam spreads, rather than an abrupt termination at $Z_{\max}$. Motivated by this behavior, the following section adopts a $3$-dB beam-depth metric based on the normalized array gain and derives the conditions under which this depth exists and remains finite.

\section{Depth-of-field of a Bessel beam}
\label{sec:dF_BB}
Following the discussion from the previous section, we define the depth-of-field as the $3$-dB beam depth along the propagation direction. Two definitions are considered: an \emph{absolute} definition referenced to the plane-wave maximum, and a \emph{peak-normalized} definition referenced to the maximum array gain achieved for the given cone angle. This section formalizes both definitions and derives the corresponding critical conditions.

\subsection{Absolute $3$-dB beam depth}
\label{subsec:absolute_3db}
The absolute definition compares the normalized gain directly with the fixed
half-power level $0.5$ (w.r.t. the plane-wave gain). For a fixed cone angle $\theta$, the
half-power distance set is $\mathcal{Z}_{0.5}^{\mathrm{abs}}(\theta)
    =
    \left\{
    z>0:
    G_{\mathrm{norm}}
    \left(
    \theta,z
    \right)
    \geq 0.5
    \right\}.$
The absolute 3-dB beam depth is then
\setcounter{equation}{14}
\begin{equation}
    \mathrm{BD}_{3\mathrm{dB}}^{\mathrm{abs}}(\theta)
    =
    \left[
    \sup \mathcal{Z}_{0.5}^{\mathrm{abs}}(\theta)
    -
    \inf \mathcal{Z}_{0.5}^{\mathrm{abs}}(\theta)
    \right].
    \label{eq:absolute_beam_depth}
\end{equation}
This definition measures whether the Bessel beam maintains at least half of the
ideal plane-wave gain. A finite absolute 3-dB beam depth requires two
conditions. 

\noindent \textit{\textbf{Condition A1:}} \textit{For the 3-dB beam depth to be finite, under the absolute definition, the $3$-dB tail must fall below $0.5$}, i.e.,
\begin{equation}
    \lim_{z\to\infty}
    G_{\mathrm{norm}}(\theta,z)
    <
    0.5.
    \label{eq:absolute_tail_condition}
\end{equation}

\noindent \textit{\textbf{Condition A2:}} \textit{For the 3-dB beam depth to exist, under the absolute definition, the axial peak must reach $0.5$}, i.e.,
\begin{equation}
    \max_{z>0}
    G_{\mathrm{norm}}(\theta,z)
    \geq
    0.5.
    \label{eq:absolute_peak_condition}
\end{equation}
\begin{figure}[!t]
    \centering
	\includegraphics[width=0.95\columnwidth]{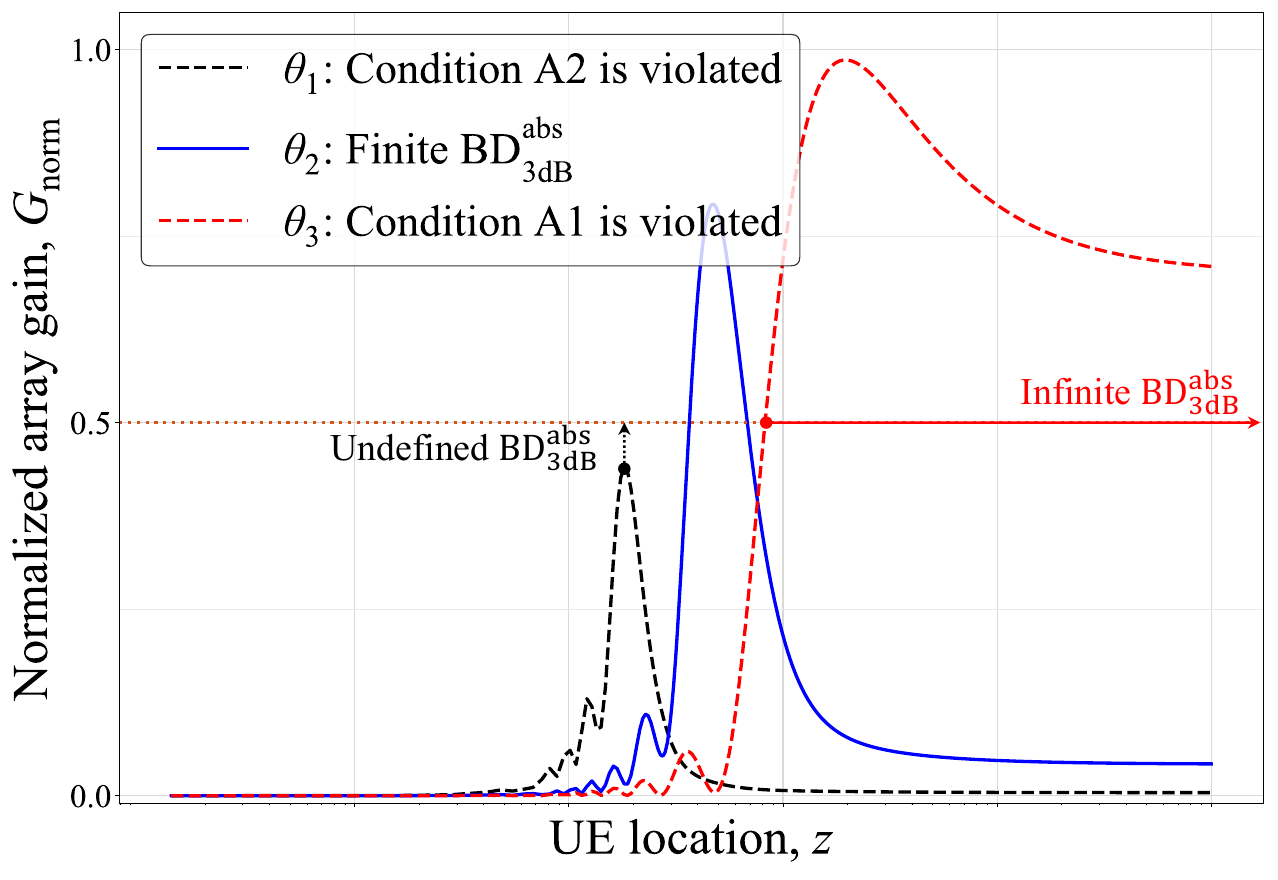}
	\caption{Illustration of the absolute $3$-dB beam depth behavior over UE location $z$. The horizontal dotted line denotes the absolute half-power threshold.}
	\label{fig:abs_illus}
    \vspace{-8pt}
\end{figure}
Fig.~\ref{fig:abs_illus} illustrates the two conditions for the absolute $3$-dB beam depth. Condition~A2 determines whether the normalized array gain reaches the half-power threshold, and therefore whether $\mathrm{BD}^{\rm abs}_{3{\rm dB}}(\theta)$ exists. If it does, Condition~A1 determines whether the corresponding region is finite or infinite. The absolute definition is useful for comparing different Bessel-beam configurations or other beamforming techniques because all gains are measured relative to the plane-wave reference.

\subsection{Peak-normalized $3$-dB beam depth}
\label{subsec:peak_normalized_3db}
The peak-normalized definition compares the gain with one half of its own axial
maximum for a given cone angle $\theta$. Hence, the corresponding half-power set is $\mathcal{Z}_{0.5}^{\mathrm{rel}}(\theta)
    =
    \left\{
    z>0:
    G_{\mathrm{norm}}
    \left(
    \theta,z
    \right)
    \geq
    0.5 \max_{z>0}
    G_{\mathrm{norm}}(\theta,z)
    \right\}.$
\begin{figure}[!t]
    \centering
	\includegraphics[width=0.95\columnwidth]{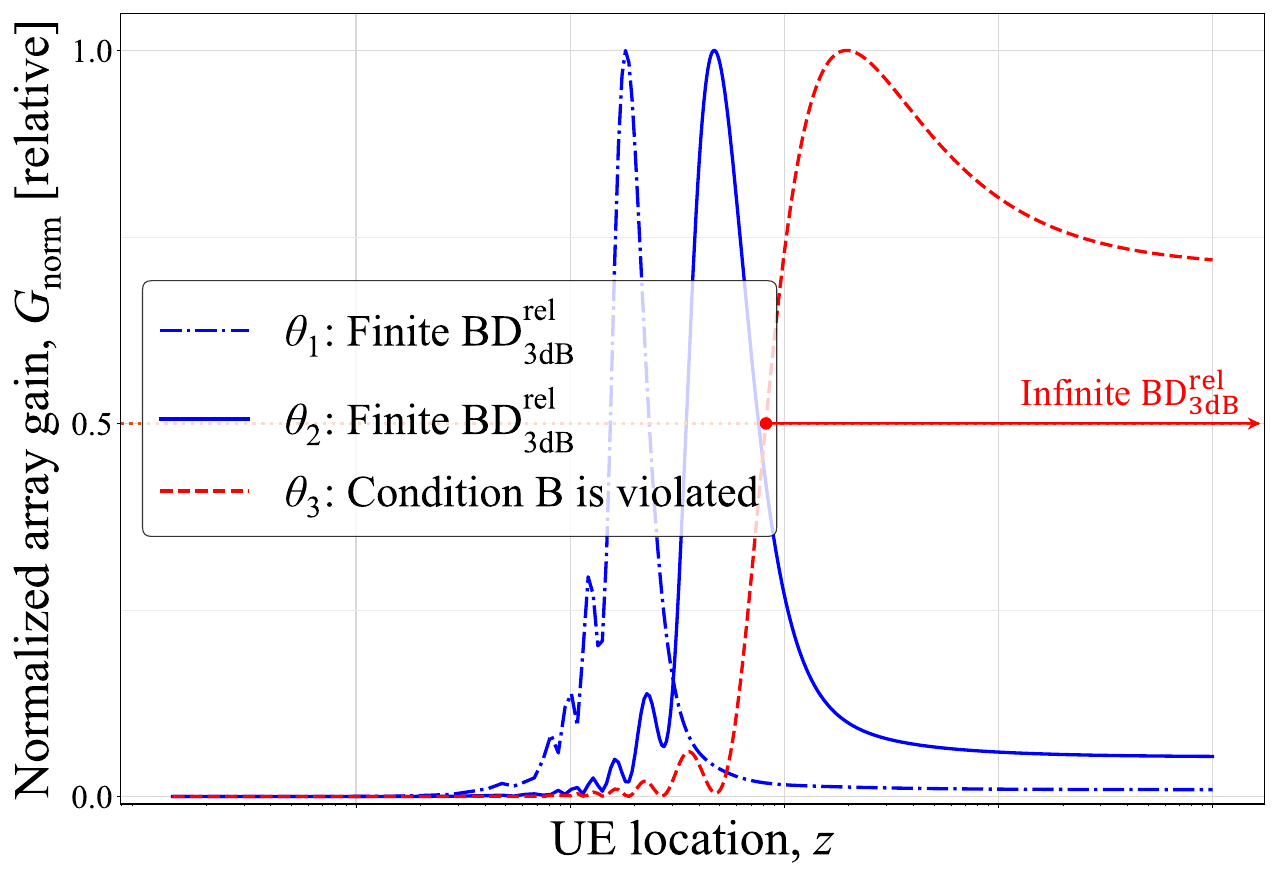}
	\caption{Illustration of peak-normalized $3$-dB beam depth behavior over UE location $z$. The horizontal dotted line denotes the relative half-power threshold.}
	\label{fig:rel_illus}
    \vspace{-8pt}
\end{figure}
The peak-normalized 3-dB beam depth is then
\begin{equation}
    \mathrm{BD}_{3\mathrm{dB}}^{\mathrm{rel}}(\theta)
    =
    \left[
    \sup \mathcal{Z}_{0.5}^{\mathrm{rel}}(\theta)
    -
    \inf \mathcal{Z}_{0.5}^{\mathrm{rel}}(\theta)
    \right].
    \label{eq:relative_beam_depth}
\end{equation}

Since the peak-normalized threshold is defined relative to the beam's own maximum, the second requirement is always satisfied (analogous to Condition A2). However, the beam depth can still become infinite, which introduces a separate condition for the peak-normalized case.

\noindent \textit{\textbf{Condition B:}} \textit{For the 3-dB beam depth to be finite, under the peak-normalized definition, we require the $3$-dB tail to fall below half of the peak. Thus, the finite-depth condition is}
\begin{equation}
    \lim_{z\to\infty}
    G_{\mathrm{norm}}(\theta,z)
    <
    \frac{1}{2}\left(\max_{z>0}
    G_{\mathrm{norm}}(\theta,z)\right).
    \label{eq:relative_tail_condition}
\end{equation}

Fig.~\ref{fig:rel_illus} illustrates the condition for the peak-normalized $3$-dB beam depth. Because the threshold is defined relative to the peak gain for each $\theta$, a half-power region always exists around the main peak. Condition~B determines whether this region is finite: if the asymptotic tail falls below the half-power threshold, the beam depth is finite; otherwise, it is infinite. The peak-normalized definition therefore measures the propagation depth of a given Bessel beam relative to its own peak gain. The absolute definition reflects communication performance relative to the plane-wave maximum, whereas the peak-normalized definition measures the propagation depth of the Bessel beam relative to its own peak.
\section{Critical Angle Characterization}
\label{sec:critical_angle_characterization}
The $3$-dB beam-depth definitions in Section~\ref{sec:dF_BB} establish critical conditions that separate the different depth-of-field regimes. In this section, we utilize these conditions to derive the corresponding critical cone angles for the \ac{cca} and \ac{csa}. These results also enable us to establish a limit on the maximum usable non-diffracting range.

\subsection{Circular array}
\label{subsec:cca_crit}
Evaluating Condition~A1 for the \ac{cca} case yields a lower bound on $\theta$ over which a Bessel beam maintains a finite depth-of-field under the absolute $3$-dB beam depth definition. Let $G_{\mathrm{norm}}^{c}(\mu_c, \infty)$ be defined as
\begin{equation}
    G_{\mathrm{norm}}^{c}(\mu_c, \infty)
    =
    \lim_{\tau_c\to\infty}
    G^{c}_{\mathrm{norm}}(\mu_c,\tau_c).
    \label{eq:ginfty_def}
\end{equation}
Using this definition we quantify the limits for a \ac{cca}: 

\noindent \textbf{Proposition 1a.} \textit{For a circular array under the absolute $3$-dB beam depth definition, $\mathrm{BD}^{\rm abs}_{\rm 3dB}$ is finite when the cone angle satisfies} 
\begin{equation}
    \theta > \sin^{-1}\left(0.5533 \frac{\lambda}{R}\right).
\end{equation}
\begin{IEEEproof}
Evaluating the limit $\tau\to\infty$ for \eqref{eq:gnorm_closed_form} (see Appendix~\ref{app:limit}) gives
\begin{equation}
    G_{\mathrm{norm}}^{c}(\mu_c, \infty)
    =
    4
    \frac{
    \xi^2+2-2\cos\xi-2\xi\sin\xi
    }{
    \xi^4
    },
    \label{eq:ginfty_real}
\end{equation}
where $\xi=\pi\mu_c/2$. For the absolute definition, the beam depth becomes infinite whenever Condition A1 is not satisfied.
Therefore, the corresponding asymptotic critical value
$\mu_c^{\mathrm{crit},A1}$ is obtained by solving the equality $G_{\mathrm{norm}}^{c}(\mu_c^{\mathrm{crit},A1},\infty)=0.5$, where the first positive solution gives $\mu_c^{\mathrm{crit},A1} = \frac{2}{\pi}\times3.47\bar{6}\approx 2.2133$ (rounding error), with the corresponding cone angle
\begin{equation}
    \theta^c_{\mathrm{crit},A1}
    =
    \sin^{-1}
    \left(
    \frac{\mu_c^{\mathrm{crit},A1}\lambda}{4R}
    \right)
    \approx
    \sin^{-1}
    \left(
    0.5533\frac{\lambda}{R}
    \right).
    \label{eq:theta_crit_2}
\end{equation}
\end{IEEEproof}
The critical cone angle associated with Condition~A1 can be obtained directly from the limits of the normalized gain in~\eqref{eq:gnorm_closed_form}, which yields $G_{\mathrm{norm}}^{c}(\mu_c, \infty)$. 
In contrast, conditions~A2 and~B depend on the maximum axial gain $\max_{z>0}G_{\mathrm{norm}}(\theta,z)$, for which obtaining an analytical expression is difficult because the stationary-point condition is implicit (see Appendix~\ref{app:implicitgmax}). Therefore, we utilize an algorithm to evaluate the peak gain numerically (see Appendix~\ref{app:peak_finder}) and utilize these values to determine the critical cone angles associated with these conditions. The accuracy of these results are presented in Section~\ref{sec:numerical_results}.

Under the absolute definition, Condition~A2 gives an upper bound on $\theta$ for a Bessel beam: if $\theta$ is too large, the peak gain remains below $0.5$, and $\mathrm{BD}^{\rm abs}_{\rm 3dB}$ is empty.

\noindent \textbf{Proposition 1b.} \textit{For a circular array under the absolute $3$-dB beam depth definition, $\mathrm{BD}^{\rm abs}_{\rm 3dB}$ is defined only when the cone angle satisfies}
\begin{equation}
    \theta \leq \sin^{-1}\left(4.056975\frac{\lambda}{R}\right).
\end{equation}
\begin{IEEEproof}
Evaluating Condition~A2 using the peak-finding algorithm gives
$\mu_c^{\mathrm{crit},A2}\approx16.2279$, with the critical cone angle:
\begin{equation}
    \theta^c_{\mathrm{crit},A2}
    =
    \sin^{-1}
    \left(
    \frac{\mu_c^{\mathrm{crit},A2}\lambda}{4R}
    \right)
    \approx
    \sin^{-1}
    \left(
    4.056975\frac{\lambda}{R}
    \right).
    \label{eq:theta_crit_1_circ}
\end{equation}
\end{IEEEproof}
We next consider the peak-normalized definition. In this case, the beam depth is always nonempty. Condition~B therefore imposes a lower bound on $\theta$, below which the peak-normalized beam depth becomes infinite.

\noindent \textbf{Proposition 2.} \textit{For a circular array under the peak-normalized $3$-dB beam depth definition, $\mathrm{BD}^{\rm rel}_{\rm 3dB}$ is finite when the cone angle satisfies }
\begin{equation}
    \theta >\sin^{-1}\left(0.5639\frac{\lambda}{R}\right).
\end{equation}
\begin{IEEEproof}
Using the peak-finding algorithm, Condition~B gives
$\mu_c^{\mathrm{crit},B}\approx2.2558$. Therefore,
\begin{equation}
    \theta_{\mathrm{crit},B}^{c}
    =
    \sin^{-1}
    \left(
    \frac{\mu^{\mathrm{crit},B}_{c}\lambda}{4R}
    \right)
    \approx
    \sin^{-1}
    \left(
    0.5639\frac{\lambda}{R}
    \right).
    \label{eq:theta_crit_3_circ}
\end{equation}
\end{IEEEproof}
Thus, Prop.~1a,~1b and~2 establish the critical cone angle thresholds that govern the existence and finiteness of both definitions of the $3$-dB beam depth for the \ac{cca} case. In the following subsection, we apply the same procedure to the \ac{csa} case and quantify the corresponding critical cone angles.
\subsection{Square array}
\label{subsec:csa_crit}
Evaluating Condition~A1 for the \ac{csa} case, similarly yields a lower bound on $\theta$ under the absolute $3$-dB beam depth definition. Let $G_{\mathrm{norm}}^{s}(\mu_s, \infty)$ be defined as
\begin{equation}
    G_{\mathrm{norm}}^{s}(\mu_s, \infty)
    =
    \lim_{\tau_s\to\infty}
    G^{s}_{\mathrm{norm}}(\mu_s,\tau_s).
    \label{eq:ginfty_sq_def}
\end{equation}
Using this definition we can quantify the limits for a \ac{csa}: 
\noindent \textbf{Proposition 3a.} \textit{For a square array under the absolute $3$-dB beam depth definition, $\mathrm{BD}^{\rm abs}_{\rm 3dB}$ is finite when the cone angle satisfies}
\begin{equation}
    \theta > \sin^{-1}\left(0.91606 \frac{\lambda}{D_{\rm AP}}\right).
\end{equation}
\begin{IEEEproof}
Evaluating the limit $\tau_s\to\infty$ for 
\eqref{eq:square_gnorm_semiclosed_form} (see Appendix~\ref{app:square_limit}) gives
\begin{equation}
\begin{aligned}
G_{\mathrm{norm}}^{s}(\mu_s, \infty)
&=
4
\left|
\int_0^{\pi/4}
\sec^2\phi
\frac{
e^{j\xi_s(\phi)}
\left[
1-j\xi_s(\phi)
\right]
-1
}{
\xi_s^2(\phi)
}
d\phi
\right|^2,
\end{aligned}
\label{eq:square_ginfty_complex_form_main}
\end{equation}
where $\xi_s(\phi)=\frac{\pi\mu_s}{2\cos(\phi)}$.
Unlike the \ac{cca} case, \eqref{eq:square_ginfty_complex_form_main}
does not simplify to a closed-form real expression because the square-aperture
boundary depends on the azimuthal direction through the $\cos(\phi)$ term.
Therefore, the asymptotic threshold must be obtained numerically using~\eqref{eq:square_ginfty_complex_form_main}. Analysis of the accuracy of the numerical evaluation is presented in Section~\ref{sec:numerical_results}. 

Similar to the \ac{cca} case, for the absolute definition, the beam depth becomes infinite whenever $G_{\mathrm{norm}}^{s}(\mu_s, \infty)\geq 0.5.$ Therefore, the corresponding asymptotic critical value
$\mu_s^{\mathrm{crit},A1}$ is obtained from $G_{\mathrm{norm}}^{s}(\mu_s^{\mathrm{crit},A1},\infty)=0.5.$
Numerically solving this equation using
\eqref{eq:square_ginfty_complex_form_main} gives the first positive solution $\mu_s^{\mathrm{crit},A1}\approx 1.83212$ (rounding error), with
the corresponding cone angle
\begin{equation}
    \theta^s_{\mathrm{crit},A1}
    =
    \sin^{-1}
    \left(
    \frac{
    \mu_s^{\mathrm{crit},A1}\lambda
    }{
    2D_{\rm AP}
    }
    \right)
    \approx
    \sin^{-1}
    \left(
    0.91606
    \frac{\lambda}{D_{\rm AP}}
    \right).
    \label{eq:square_theta_crit_2_Dap}
\end{equation}
\end{IEEEproof}
Similar to the \ac{cca} case the max gain cannot be calculated analytically hence we utilize the peak-finding algorithm to quantify the critical cone angle associated with Condition~A2, and~B. Evaluating Condition~A2 for a \ac{csa} similarly, gives an upper bound on $\theta$:

\noindent \textbf{Proposition 3b.} \textit{For a square array under the absolute $3$-dB beam depth definition, $\mathrm{BD}^{\rm abs}_{\rm 3dB}$ is defined only when the cone angle satisfies} 
\begin{equation}
    \theta\leq\sin^{-1}\left(4.99875\frac{\lambda}{D_{\rm AP}}\right).
\end{equation}
\begin{IEEEproof}
Numerically solving Condition~A2 using the peak-finding algorithm for the \ac{csa} case gives
$\mu_s^{\mathrm{crit},A2}\approx9.9975$, with the critical cone angle:
\begin{equation}
    \theta^s_{\mathrm{crit},A2}
    =
    \sin^{-1}
    \left(
    \frac{\mu_s^{\mathrm{crit},A2}\lambda}{2D_{\rm AP}}
    \right)
    \approx
    \sin^{-1}
    \left(
    4.99875\frac{\lambda}{D_{\rm AP}}
    \right).
    \label{eq:theta_crit_1_sq}
\end{equation}
\end{IEEEproof}

Finally, evaluating Condition~B for a \ac{csa} gives a lower bound on $\theta$ for a Bessel beam under the peak-normalized definition, giving:

\noindent \textbf{Proposition 4.} \textit{For a square array under the peak-normalized $3$-dB beam depth definition, $\mathrm{BD}^{\rm rel}_{\rm 3dB}$ is finite when the cone angle satisfies}
\begin{equation}
    \theta >\sin^{-1}\left(0.9383\frac{\lambda}{D_{\rm AP}}\right).
\end{equation}
\begin{IEEEproof}
Using the peak-finding algorithm, numerical calculation of Condition~B gives
$\mu_s^{\mathrm{crit},B}\approx1.8765$. The corresponding cone angle is
\begin{equation}
    \theta_{\mathrm{crit},B}^{s}
    =
    \sin^{-1}
    \left(
    \frac{\mu^{\mathrm{crit},B}_{s}\lambda}{2D_{\rm AP}}
    \right)
    \approx
    \sin^{-1}
    \left(
    0.9383\frac{\lambda}{D_{\rm AP}}
    \right).
    \label{eq:theta_crit_3_sq}
\end{equation}
\end{IEEEproof}
Thus, Prop.~3a,~3b and~4 establish the critical cone angle thresholds for the \ac{csa} case under the absolute and peak-normalized $3$-dB beam-depth definitions. Together with the \ac{cca} results in Subsection~\ref{subsec:cca_crit}, these
thresholds show that a phase-based Bessel beam is usable only over a restricted cone angle interval. In the following
subsection, we utilize the \ac{cca} and \ac{csa} results and define the operating ranges.
\subsection{Usable range of Bessel beams}
\label{subsec:Summary}
The results in Subsections~\ref{subsec:cca_crit} and~\ref{subsec:csa_crit} define the usable cone angle ranges for phase-based Bessel beams under the absolute and peak-normalized beam depth definitions. For the \ac{cca}, the absolute $3$-dB beam depth is nonempty and finite only when both Conditions~A1 and~A2 are satisfied. Hence, the usable cone angle range is a result of Prop. 1a and 1b,
\begin{equation}
    \sin^{-1}\left(0.5533\frac{\lambda}{R}\right)
    <
    \theta
    \leq
    \sin^{-1}\left(4.056975\frac{\lambda}{R}\right).
    \label{eq:usable_theta_abs_circ}
\end{equation}
Similarly, for the \ac{csa} case, the corresponding usable cone angle range is a result of Prop. 3a and 3b,
\begin{equation}
    \sin^{-1}\left(0.91606\frac{\lambda}{D_{\rm AP}}\right)
    <
    \theta
    \leq
    \sin^{-1}\left(4.99875\frac{\lambda}{D_{\rm AP}}\right).
    \label{eq:usable_theta_abs_sq}
\end{equation}
In both cases, the lower bound on $\theta$ corresponds to Condition~A1, while the upper bound corresponds to Condition~A2.
For the peak-normalized definition, the half-power region is always nonempty, but it is finite only when Condition~B is satisfied. Prop. 2 and Prop. 4 therefore define the corresponding usable ranges for the \ac{cca} and \ac{csa}, respectively.

From Prop.~1a, 2, 3a, and~4, we notice that the critical cone angles are proportional to $\lambda/R$ or $\lambda/D_{\rm AP}$. Since practical communication systems typically satisfy $\lambda\ll R$ and $\lambda\ll D_{\rm AP}$, i.e., being electrically large, the corresponding critical cone angles, $\theta^{c}_{\rm crit,A1}$, $\theta^{c}_{\rm crit,B}$, $\theta^{s}_{\rm crit,A1}$, and $\theta^{s}_{\rm crit,B}$, lie in the small-angle regime. Consequently, when the approximation $\tan(\theta)\approx\sin(\theta)$ holds, the maximum usable non-diffracting range can be expressed as $Z^{\rm c}_{\max}\approx d^c_{\mathrm{FA}}/(2\mu_c)$ for the \ac{cca} and $Z^{\rm s}_{\max}\approx d^s_{\mathrm{FA}}/(2\sqrt{2}\mu_s)$ for the \ac{csa}. The resulting critical values of $Z_{\max}$ are summarized in Table~\ref{tab:zmax_crit}.
\begin{table}[t]
    \centering
    \caption{Critical geometric non-diffracting ranges}
    \label{tab:zmax_crit}
    \begin{tabular}{c c}
        \toprule
        \textbf{Proposition} & \textbf{Critical $Z_{\max}$} \\
        \midrule
        Proposition~1a
        &
        $\displaystyle
        \frac{R}
        {\tan\left(\sin^{-1}\left(0.5533\frac{\lambda}{R}\right)\right)}
        \approx
        0.226\,d_{\rm FA}^{c}
        $
        \\[4ex]

        Proposition~2
        &
        $\displaystyle
        \frac{R}
        {\tan\left(\sin^{-1}\left(0.5639\frac{\lambda}{R}\right)\right)}
        \approx
        0.222\,d_{\rm FA}^{c}
        $
        \\[4ex]

        Proposition~3a
        &
        $\displaystyle
        \frac{D_{\rm AP}}
        {\sqrt{2}\tan\left(\sin^{-1}\left(0.91606\frac{\lambda}{D_{\rm AP}}\right)\right)}
        \approx
        0.193\,d_{\rm FA}^{s}
        $
        \\[4ex]

        Proposition~4
        &
        $\displaystyle
        \frac{D_{\rm AP}}
        {\sqrt{2}\tan\left(\sin^{-1}\left(0.9383\frac{\lambda}{D_{\rm AP}}\right)\right)}
        \approx
        0.188\,d_{\rm FA}^{s}
        $
        \\
        \bottomrule
    \end{tabular}
    \vspace{-8pt}
\end{table}

These results show that both beam depth definitions impose a finite upper limit on the usable non-diffracting range. For the circular aperture, this limit is approximately $0.22d^c_{\rm FA}$ under both definitions. For the square aperture, the corresponding limit is approximately $0.19d^s_{\rm FA}$. The close agreement between these limits arises because the peak normalized array gain remains close to the plane-wave reference near the critical cone angle. Consequently, the absolute threshold $0.5$ and the peak-normalized threshold $0.5G_{\rm peak}$ are similar, causing the two beam-depth definitions to yield nearly identical limiting ranges. Thus, although the absolute limiting range scales with array size and wavelength through $d_{\rm FA}$, its normalized value approaches a geometry-dependent constant.

\section{Numerical Analysis}
\label{sec:numerical_results}
In this section, we first verify the accuracy of the Fresnel approximations underlying the normalized array-gain expressions in~\eqref{eq:gnorm_closed_form} and~\eqref{eq:square_gnorm_semiclosed_form} for the \ac{cca} and \ac{csa}, respectively. We then validate the critical cone angles obtained from Conditions~A1, A2, and~B which rely on the convergence of the numerical integration for the \ac{csa} and the accuracy of the peak-finding algorithm (discussed in Appendices~\ref{app:csa_nphi_accuracy} and~\ref{app:peak_finder}, respectively). 

Building on these validations, we compare the analytical maximum usable non-diffracting range with the numerically obtained $3$-dB beam depth, and then show the impact of discretization on the derived analytical limits. Finally, we investigate the engineering implications through comparisons between \ac{cca} and \ac{csa}, and practical $Z_{\max}$ limits for future near-field systems at varying array sizes and carrier frequencies.

\subsection{Accuracy analysis}
We first validate the Fresnel approximations used to derive the normalized array-gain expressions in~\eqref{eq:gnorm_closed_form} and~\eqref{eq:square_gnorm_semiclosed_form}. As shown in Sections~\ref{subsec:CCA_gnorm} and~\ref{subsec:CSA_gnorm}, the approximation error remains below $3.5\times10^{-3}$ provided that the \ac{ue} distance satisfies $z\geq2.5R$ for the \ac{cca} and $z\geq1.2D_{\rm AP}$ for the \ac{csa}. Restricting the analysis to the radiative near-field reduces these conditions to $R>2\lambda$ and $D_{\rm AP}>1.32\lambda$, respectively, which are readily satisfied for electrically large apertures.

\begin{figure}[t!]
\centering

\subfloat[Comparison of~\eqref{eq:normalized_array_gain_general} with~\eqref{eq:gnorm_closed_form} for a circular array]{%
\includegraphics[width=0.97\columnwidth]{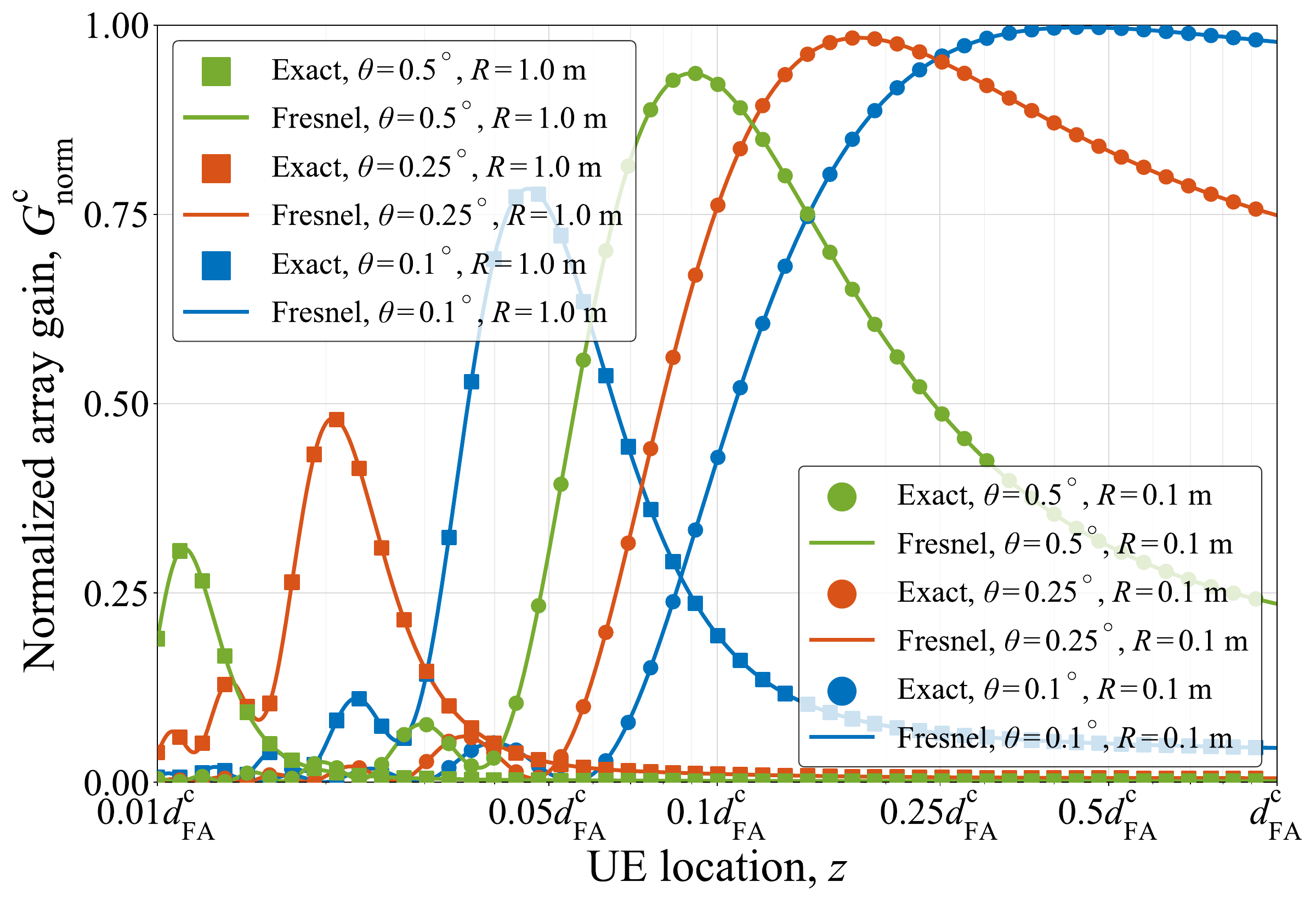}
\label{fig:G_norm_circ_cform_vs_Exact}}
\hfill

\subfloat[Comparison of~\eqref{eq:normalized_array_gain_general} with~\eqref{eq:square_gnorm_semiclosed_form} for a square array]{%
\includegraphics[width=0.97\columnwidth]{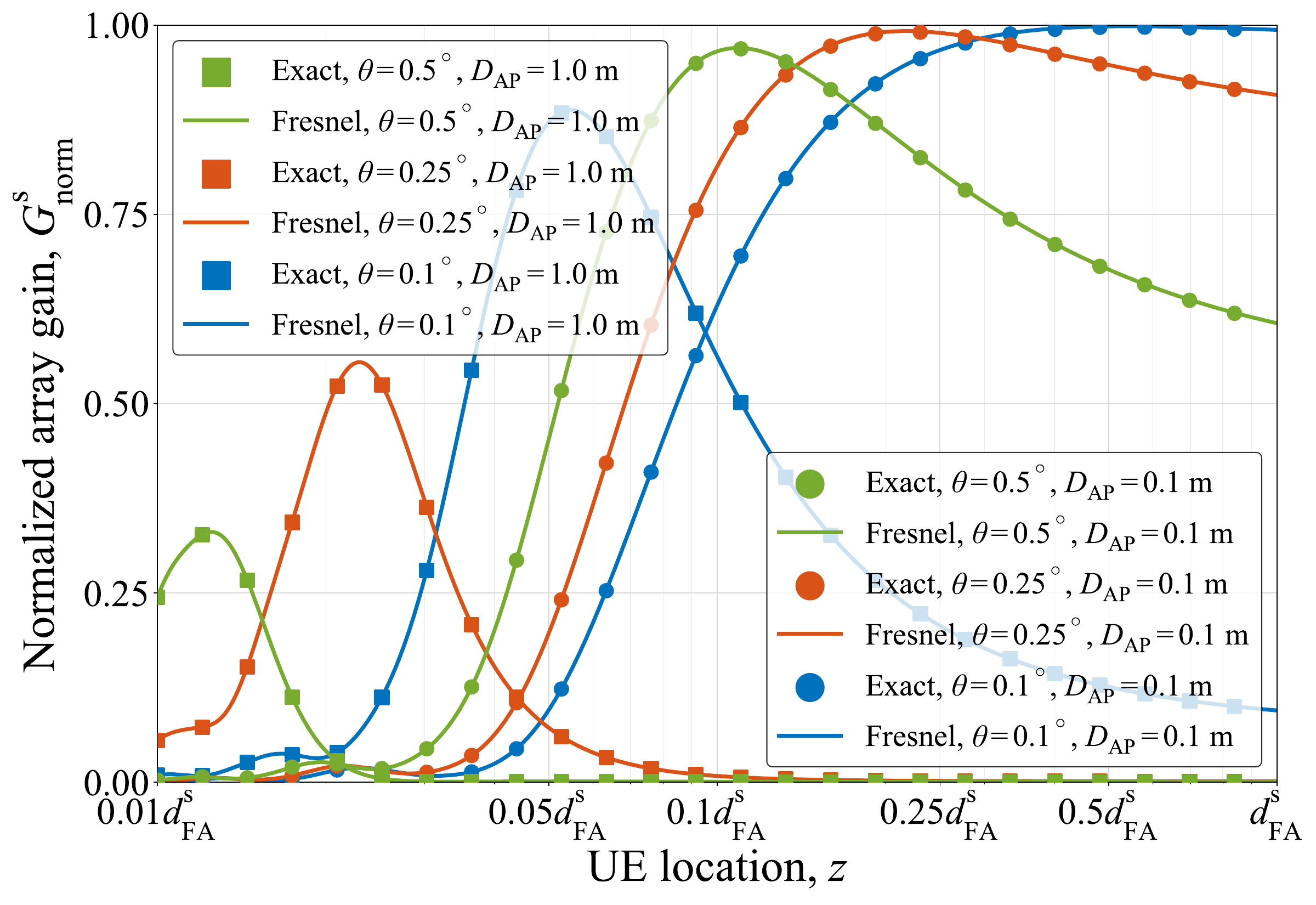}
\label{fig:G_norm_sqar_cform_vs_Exact}}

\caption{Fresnel approximation compared with the exact integral $G_{\rm norm}$.}
\label{fig:exact_vs_fresnel}
\vspace{-14pt}
\end{figure}

Figs.~\ref{fig:G_norm_circ_cform_vs_Exact} and~\ref{fig:G_norm_sqar_cform_vs_Exact} compare the exact normalized array gain with the proposed Fresnel-based expressions for representative array sizes ($R,D_{\rm AP}=\{0.1,1.0\}$~m), cone angles ($\theta=\{0.5^\circ,0.25^\circ,0.1^\circ\}$), and a carrier frequency of $300$~GHz. For both array geometries, the proposed expressions closely match the exact solution. This validates the Fresnel-based expressions underlying the analytical evaluation of Condition~A1 and the corresponding critical cone-angle thresholds. Conditions~A2 and~B additionally depend on the numerically evaluated peak array gain. The peak-finding algorithm used for their evaluation is validated in Appendix~\ref{app:peak_finder}, with relative errors in the critical angles below $0.01\%$ for both the \ac{cca} and \ac{csa}.

\begin{figure*}[t!]
    \centering
    \subfloat[Conditions A1 and A2]{%
        \includegraphics[width=0.98\columnwidth]{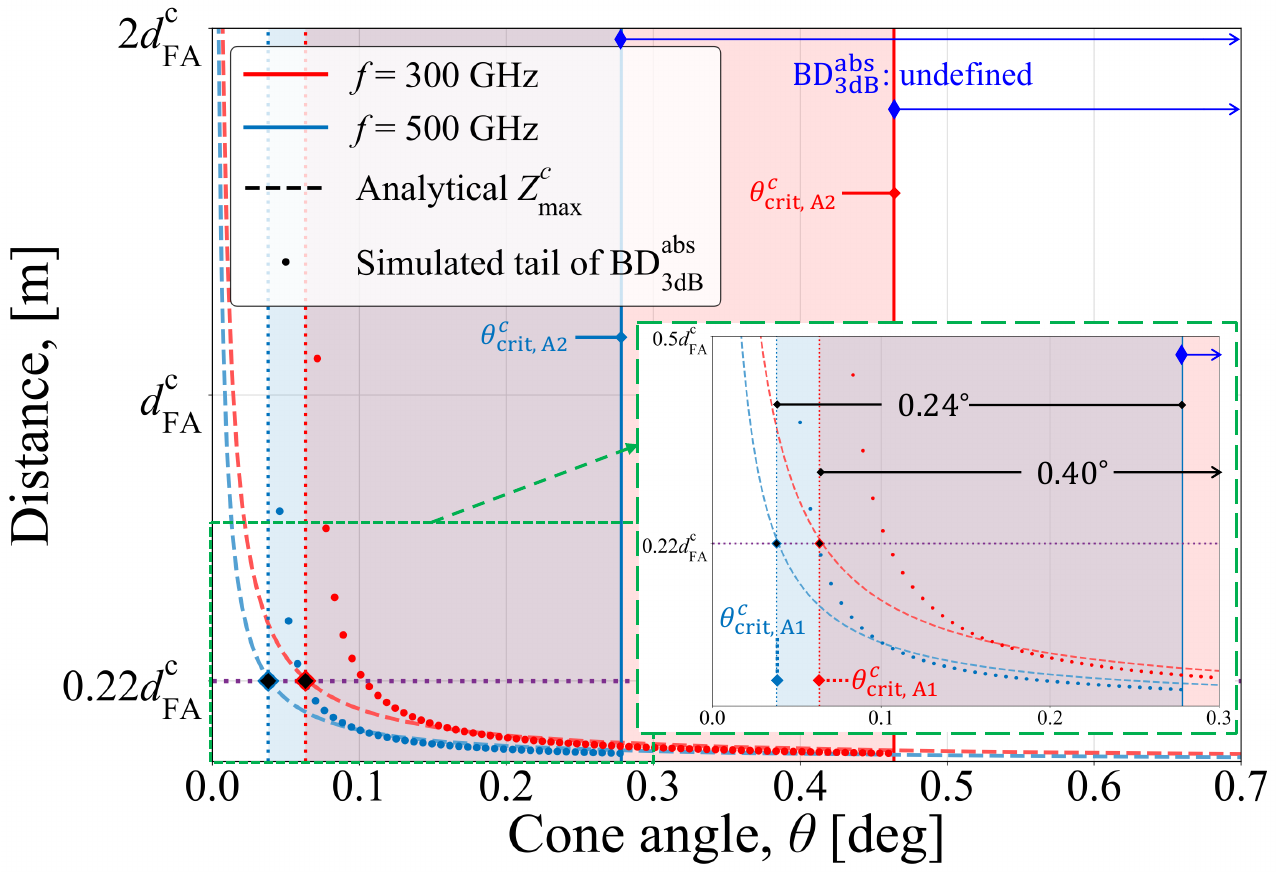}
        \label{fig:cca_conditions_a1_a2}
    }
    \hfill
    \subfloat[Condition B]{%
        \includegraphics[width=0.98\columnwidth]{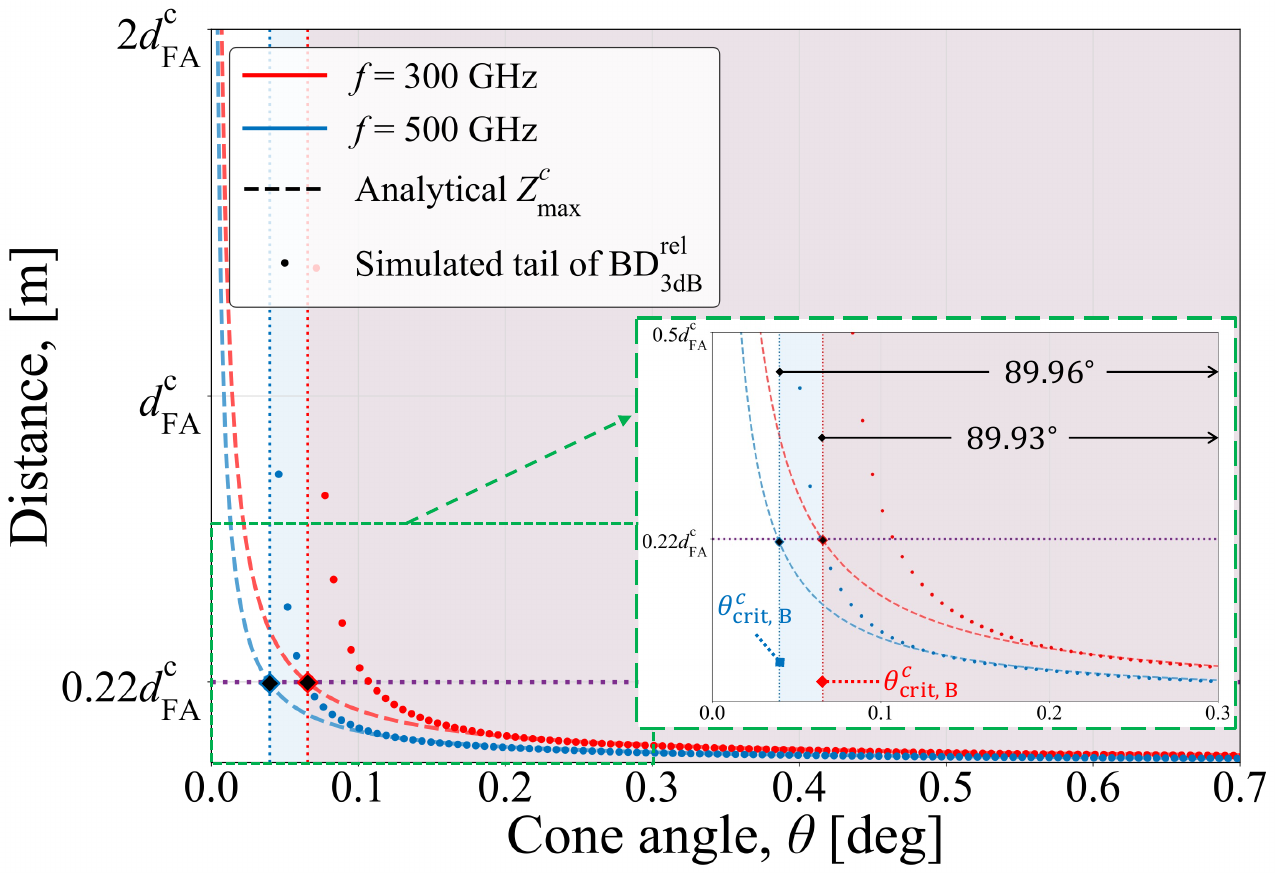}
        \label{fig:cca_condition_b}
    }
    \caption{Comparison between the geometric non-diffracting range \(Z_{\max}\) and
    the 3-dB beam-depth-based usable range for the \ac{cca} at $R=0.5$~m.
    (a) Conditions~A1 and A2 based on the absolute 3-dB beam depth.
    (b) Condition~B based on the peak-normalized 3-dB beam depth.}
    \label{fig:cca_conditions}
    \vspace{-14pt}
\end{figure*}
\begin{figure*}[t!]
    \centering
    \subfloat[Conditions A1 and A2]{%
        \includegraphics[width=0.98\columnwidth]{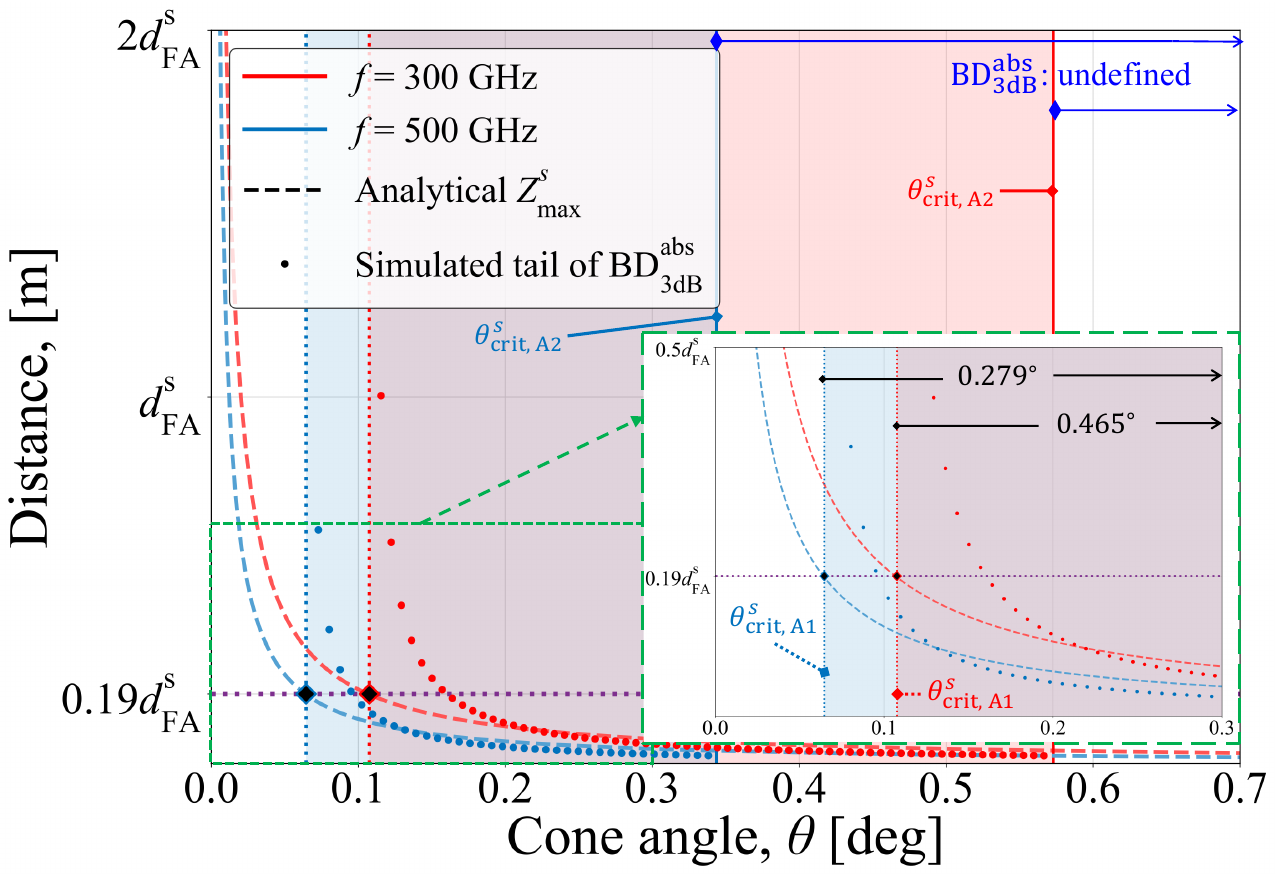}
        \label{fig:csa_conditions_a1_a2}
    }
    \hfill
    \subfloat[Condition B]{%
        \includegraphics[width=0.98\columnwidth]{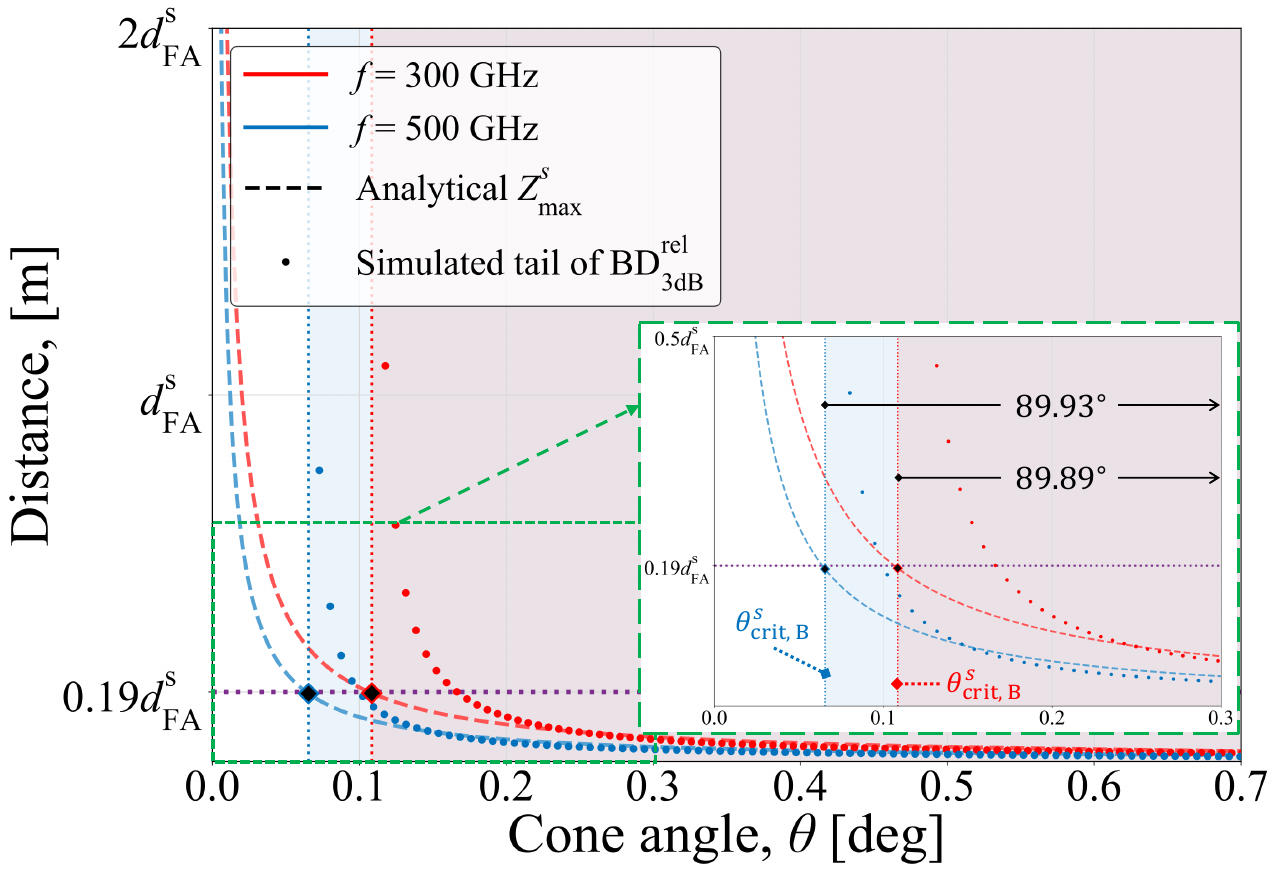}
        \label{fig:csa_condition_b}
    }
    \caption{Comparison between the geometric non-diffracting range \(Z_{\max}\) and
    the 3-dB beam-depth-based usable range for the \ac{csa} at $D_{\rm AP}=0.5$~m.
    (a) Conditions~A1 and A2 based on the absolute 3-dB beam depth.
    (b) Condition~B based on the peak-normalized 3-dB beam depth.}
    \label{fig:csa_conditions}
    \vspace{-10pt}
\end{figure*}
\subsection{Critical $Z_{\max}$ and the 3-dB beam depth}
\label{subsec:zmax_beam_depth_relationship}
In this subsection, we compare the geometric non-diffracting range $Z_{\max}$ with the corresponding 3-dB beam depth. As discussed in Section~\ref{sec:dF_BB}, $Z_{\max}$ provides a geometric estimate of the finite-aperture non-diffracting range but does not capture the evolution of the beam's spatial confinement, which is instead characterized by the 3-dB beam depth. Figs.~\ref{fig:cca_conditions} and~\ref{fig:csa_conditions} compare the analytical $Z_{\max}$ (dashed curves) with the numerically obtained 3-dB tail location (markers) for the \ac{cca} and \ac{csa}, respectively, at $300$ and $500$~GHz, with the insets highlighting the critical cone angles. We consider smaller angles to highlight the limiting cone angle regime.

For Conditions~A1 and~A2 (Figs.~\ref{fig:cca_conditions_a1_a2} and~\ref{fig:csa_conditions_a1_a2}), a finite absolute 3-dB beam depth exists only for
$\theta_{\rm crit,A1}<\theta\leq\theta_{\rm crit,A2}$.
As $\theta\rightarrow\theta_{\rm crit,A1}$, the beam depth diverges, marking the onset of the infinite depth regime predicted by Prop.~1a and~3a, whereas beyond $\theta_{\rm crit,A2}$ it becomes undefined because the normalized array gain no longer exceeds the absolute 3-dB threshold. The limits of $0.22d_{\rm FA}^{c}$ and $0.19d_{\rm FA}^{s}$ intersect $Z_{\max}$ almost exactly at $\theta_{\rm crit,A1}$, visually confirming the derived maximum usable non-diffracting ranges. Increasing the operating frequency shifts both A1 and A2 thresholds toward smaller angles and reduces the absolute angular width of the finite-depth interval.

For the peak-normalized criterion (Figs.~\ref{fig:cca_condition_b} and~\ref{fig:csa_condition_b}), the 3-dB beam depth remains defined for all cone angles, with $\theta_{\rm crit,B}$ marking the transition from finite to infinite depth. Its intersection with $Z_{\max}$ again coincides closely with the corresponding asymptotic limit, confirming the predicted critical range. Increasing frequency lowers the single threshold $\theta_{\rm crit,B}$, thereby enlarging the set of cone angles satisfying the finite-depth condition. Finally, $\theta_{\rm crit,B}$ lies close to $\theta_{\rm crit,A1}$ for both array geometries, showing that the two criteria identify nearly the same onset of infinite depth behavior.
\begin{figure*}[t]
\setcounter{figure}{8}
    \centering
    \subfloat[Illustration of array geometries]{%
        \includegraphics[width=0.55\columnwidth]{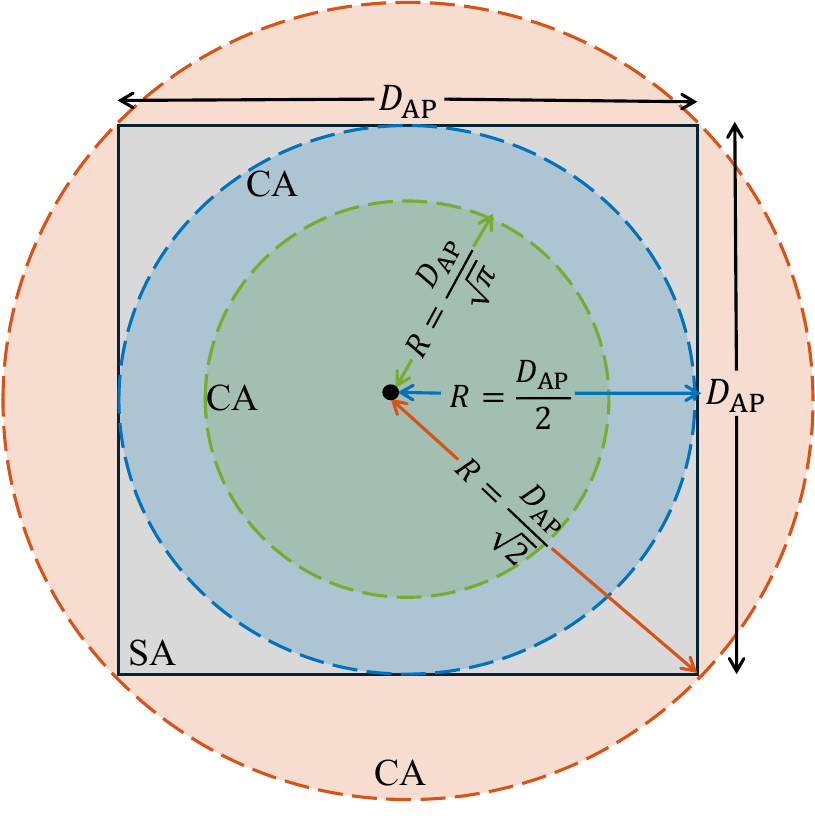}%
        \label{fig:illus}
    }
    \hfill
    \subfloat[Critical cone angle (Condition~A1 and~B)]{%
        \includegraphics[width=0.66\columnwidth]{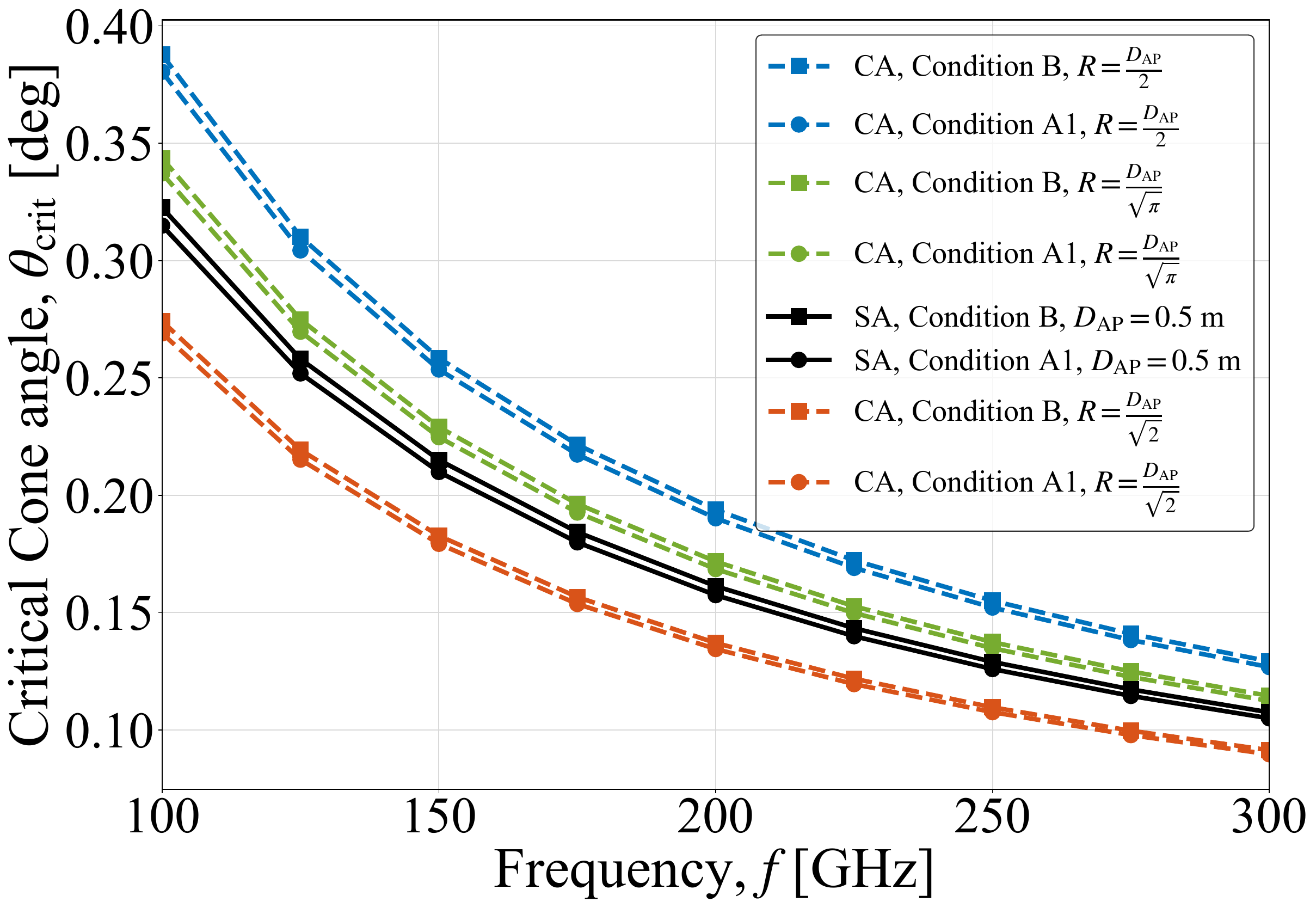}%
    \label{fig:_freq}
    }
    \hfill
    \subfloat[Critical non-diffracting range]{%
        \includegraphics[width=0.66\columnwidth]{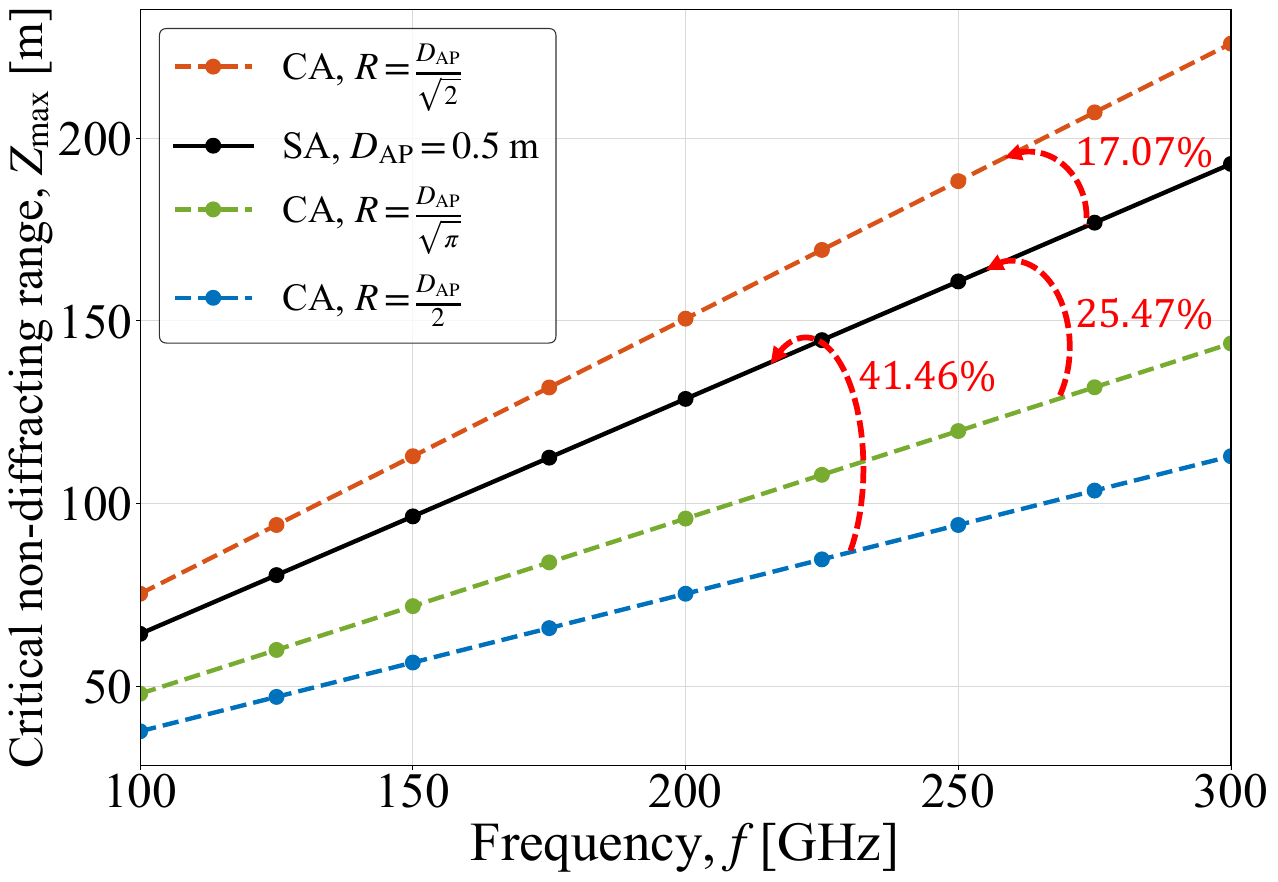}%
    \label{fig:array}
    }
    \caption{Impact of array geometry on the critical cone angle and maximum usable non-diffracting range for \ac{cca} and \ac{csa} under different aperture criteria.}
\label{fig:critical_theta_frequency_CCA_CSA}
\vspace{-8pt}
\end{figure*}

\begin{figure}[!t]
\setcounter{figure}{7}
    \centering
    \subfloat[Discrete circular array]{%
        \includegraphics[width=0.98\columnwidth]{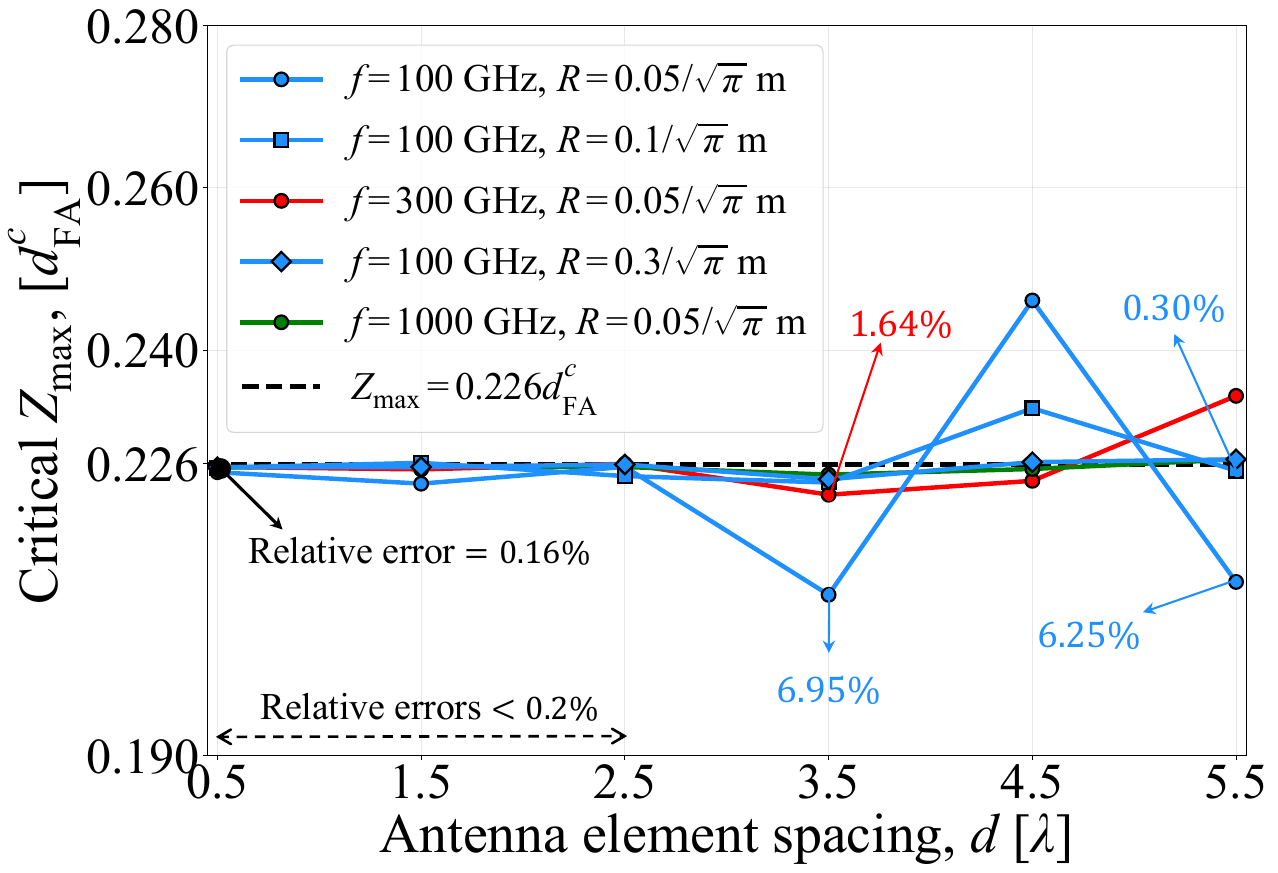}%
        \label{fig:disc_circ}
    }
    \hfill
    \subfloat[Discrete square array]{%
        \includegraphics[width=0.98\columnwidth]{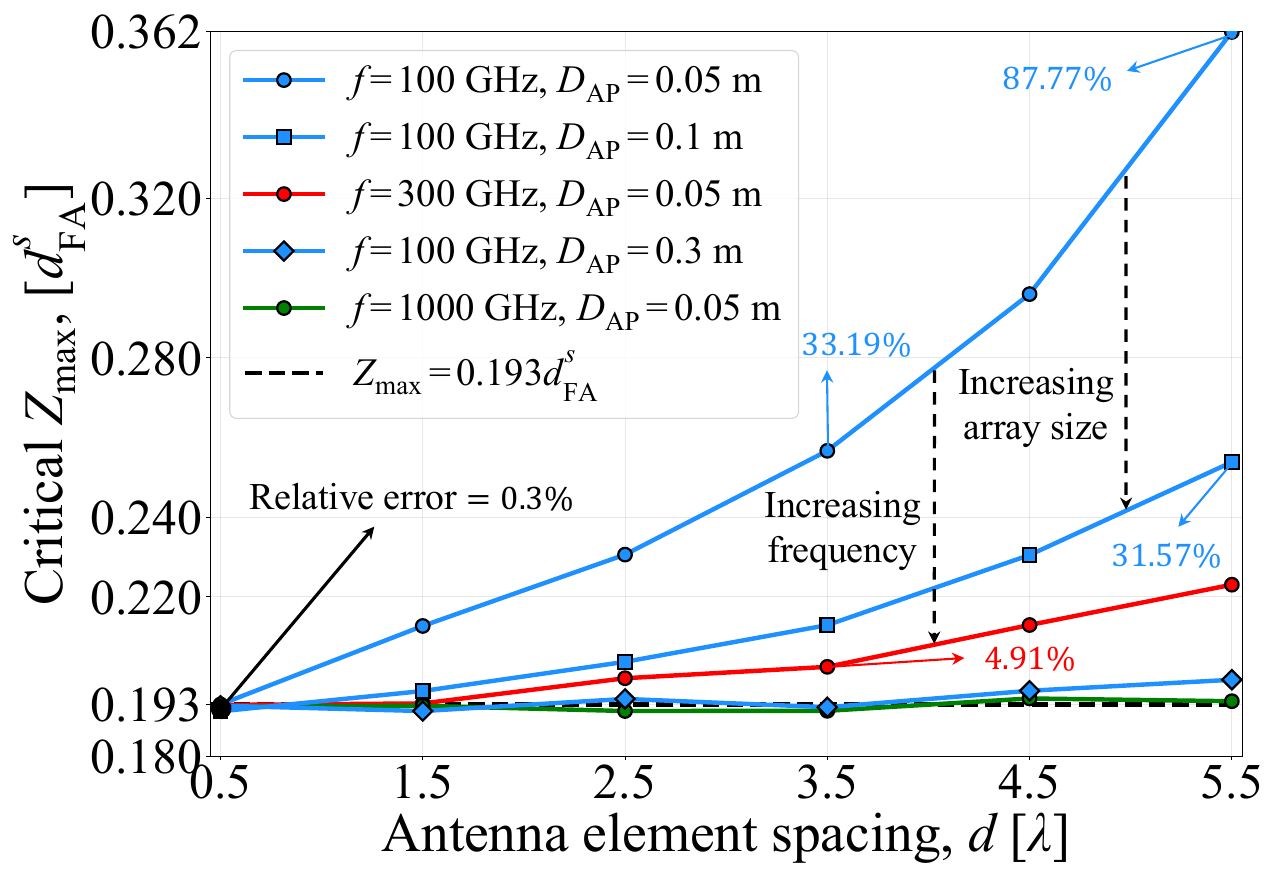}%
    \label{fig:disc_square}
    }
    \caption{Critical $Z_{\max}$ obtained numerically and analytically (under continuous array approximation) for discrete arrays at varying antenna element spacing.}
    \label{fig:dis_array}
    \vspace{-14pt}
\end{figure}

\subsection{Effect of discretization}
\label{subsec:discrete_array_valid}
For practical discrete array applications, we now evaluate the derived array limits numerically across different antenna element spacings, carrier frequencies, and array sizes. Fig.~\ref{fig:disc_circ} compares the critical non-diffracting range of the discrete circular array (antenna elements arranged in a square grid) with the analytical limit of $0.22d^{c}_{FA}$ from Prop.~1a for spacings from $0.5\lambda$ to $5.5\lambda$. At half-wavelength spacing, the maximum relative error is only $0.16\%$, remaining below $0.2\%$ up to $\approx 2.5\lambda$ and below $7\%$ even for the sparsest array.

Fig.~\ref{fig:disc_square} shows the corresponding results for the discrete square array. At half-wavelength spacing, the maximum relative error remains approximately $0.3\%$; however, the deviation increases with spacing, particularly for larger apertures and lower frequencies, reaching $87.8\%$ for the sparsest configurations. Increasing the operating frequency improves the accuracy, with errors below $\approx 5\%$ even at the largest spacings.

Thus, the limits derived using continuous array approximation accurately predict both array geometries for half-wavelength discrete arrays. The critical non-diffracting range prediction for the circular array remains robust under sparse sampling, whereas the square array is more sensitive to discretization. For both geometries, agreement improves with increasing frequency and array size as the number of elements increases and the discrete aperture more closely resemble the continuous array approximation.

\subsection{Effect of array geometry}
Having established the critical cone angles and maximum usable non-diffracting ranges for the \ac{cca} and \ac{csa}, we next examine the impact of array geometry on these limits in Fig.~\ref{fig:critical_theta_frequency_CCA_CSA}. Since a square aperture has no unique equivalent circular radius, we consider three common equivalence criteria: the inscribed ($R=D_{\rm AP}/2$), equal-area ($R=D_{\rm AP}/\sqrt{\pi}$), and circumscribed ($R=D_{\rm AP}/\sqrt{2}$) circles, all illustrated in Fig.~\ref{fig:illus}.

Fig.~\ref{fig:_freq} compares the resulting critical cone angles for Conditions~A1 and~B. The critical angles decrease with frequency and remain nearly identical between the two conditions. The inscribed and circumscribed circles provide the upper and lower bounds, respectively, while the \ac{csa} lies between them and is most closely approximated by the equal-area circle.

The corresponding limiting $Z_{\max}$ values in Fig.~\ref{fig:array} follow the same geometric ordering, with the inscribed and circumscribed circles providing the lower and upper bounds, respectively. The \ac{csa} limit is approximately $17.1\%$ larger than that of the equal-area circle, while the circumscribed-circle limit is approximately $25.5\%$ larger than the \ac{csa}; the two circular bounds differ by more than $41.5\%$. Thus, the limiting non-diffracting range depends on array geometry, with the circular cases providing bounds for the \ac{csa} and the equal-area circle giving the closest approximation.

\begin{figure*}[!t]
\setcounter{figure}{9}
    \centering
    \subfloat[Half-wavelength spaced circular arrays]{%
        \includegraphics[width=0.99\columnwidth]{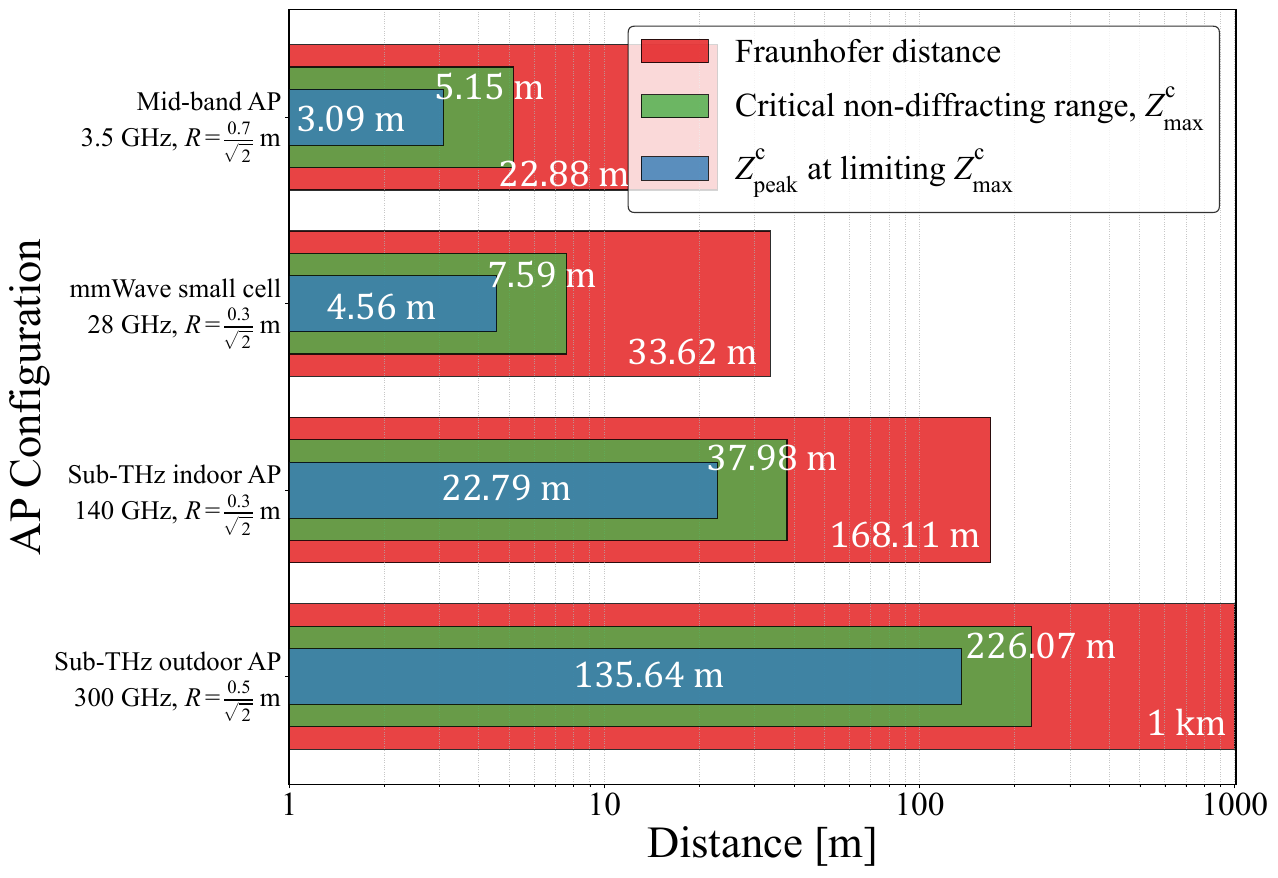}
        \label{fig:practical_aperture_sizes_A}
    }
    \hfill
    \subfloat[Half-wavelength spaced square arrays]{%
        \includegraphics[width=0.99\columnwidth]{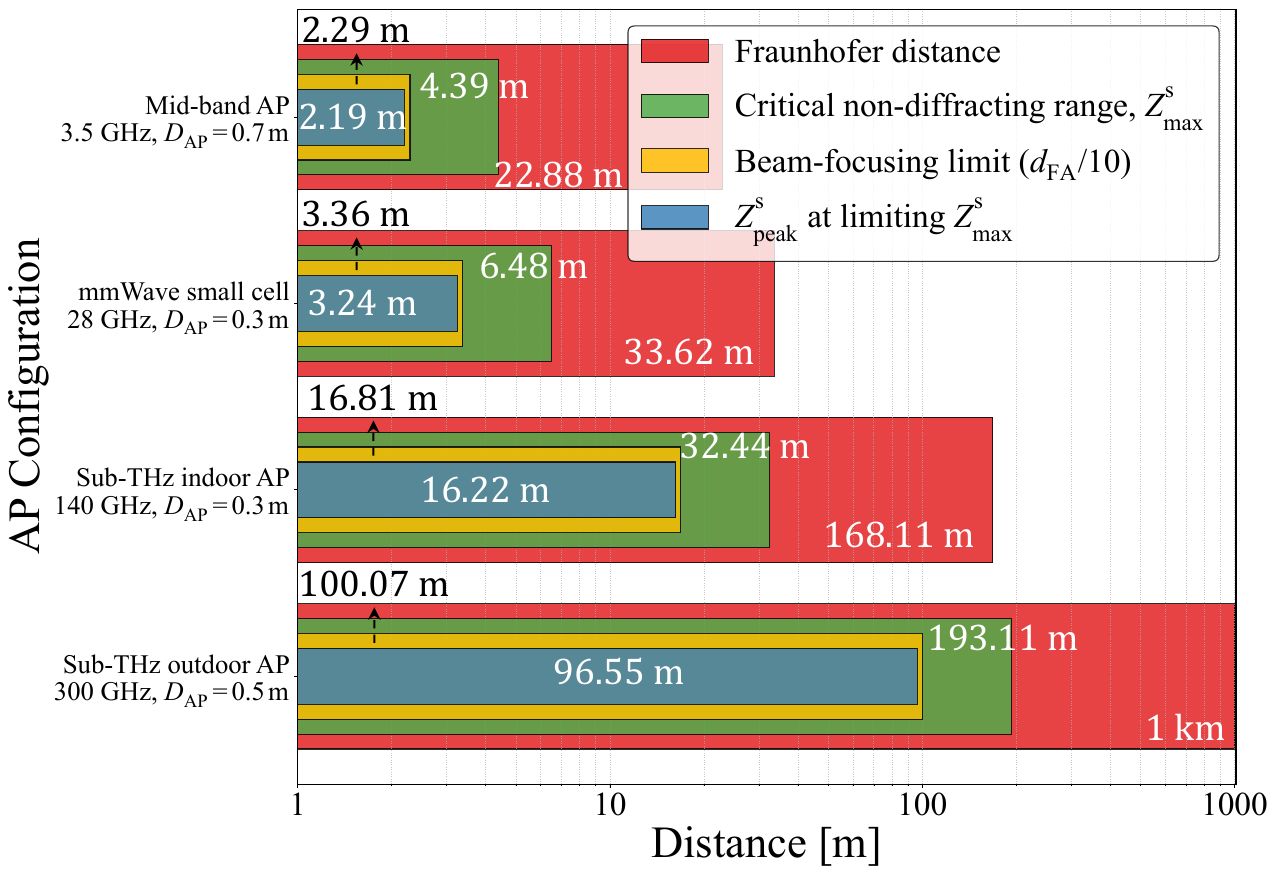}
        \label{fig:practical_aperture_sizes_B}
    }
    \caption{Comparing the conventional Fraunhofer distance, maximum $Z_{\max}$ of a Bessel beam with a finite depth-of-field, and the corresponding peak location.}
    \label{fig:practical_aperture_sizes}
\vspace{-14pt}
\end{figure*}
\subsection{Implications for future near-field communication systems}
We now illustrate the implications of the derived limits for different microwave, mmWave, and sub-THz transmitter configurations in Fig.~\ref{fig:practical_aperture_sizes}. To explore the fundamental limits numerically, we deliberately range the configurations from the microwave $3.5$\,GHz configuration all the way to futuristic large-scale sub-THz arrays operating at $300$\,GHz. Specifically, Fig.~\ref{fig:practical_aperture_sizes_A} presents the results for the CA, while Fig.~\ref{fig:practical_aperture_sizes_B} focuses on SA metrics. To be consistent with previous results, all cases employ half-wavelength-spaced discrete arrays ($d=\lambda/2$), with \ac{cca} and \ac{csa} having the same $d_{\rm FA}$ (i.e. $R= D_{\rm AP}/\sqrt{2}$). 

Fig.~\ref{fig:practical_aperture_sizes_A} illustrates the Fraunhofer distance, maximum usable non-diffracting range, and corresponding axial peak location for different \acp{cca}. Although the maximum non-diffracting range with a finite depth-of-field occupies only a fraction of the Fraunhofer distance, it can still remain non-negligible for future near-field communication systems.

For instance, at $140$~GHz, a circular indoor \ac{ap} with $R=0.3/\sqrt{2}$~m $\approx$ $21$\,cm achieves a critical $Z^{c}_{\max}$ of approximately $38$~m, with an axial peak near $\approx 23$~m. Similar observations hold for the corresponding \acp{csa} in Fig.~\ref{fig:practical_aperture_sizes_B}. Furthermore, for the same Fraunhofer distance, \acp{csa} exhibit slightly shorter maximum non-diffracting ranges than their circular counterparts, reaching approximately $32$~m at $140$~GHz and $193$~m at $300$~GHz with a $0.5$$\times$$0.5$\,m transmitter array. Importantly, the established limit do not preclude the practical use of Bessel beams beyond those, but rather establish the maximum propagation range over which their finite-depth behavior can be exploited. Finally, the results in Fig.~\ref{fig:practical_aperture_sizes_B} present a close match between the peak power locations at the finite depth-of-field boundary for the focused beam and the Bessel beam, indicating that there could be a similar general limit for \emph{any} type of near-field beam to be explored in the future work.

\section{Conclusion}
\label{sec:conclusion}
In this work, we analytically characterized the finite-depth behavior of phase-based Bessel beams for near-field communications. First, we derived expressions for the normalized array gain of an ideal \ac{cca} and a \ac{csa}. We then derived absolute and peak-normalized $3$-dB depth-of-field criteria for both circular and square geometries. Later, from the principal conditions governing the existence and finiteness of the Bessel beam depth in different setups, we obtained critical cone-angle thresholds that partitioned the Bessel beam configuration space into: (i)~undefined, (ii)~finite, and (iii)~infinite depth-of-field regimes. Finally, these thresholds enabled us to determine the maximum usable non-diffracting range of a Bessel beam generated by a given array configuration.

The analysis shows that the conventional non-diffracting range $Z_{\max}$ alone does \emph{not} determine whether a Bessel beam maintains a finite or infinite depth of field. The actual finite depth-of-field range depends on the critical Bessel cone angle, which is subject to: (i)~the array aperture, (ii)~the array shape, and (iii)~the signal frequency. Importantly, for electrically large antennas, this leads to a maximum usable non-diffracting range of $\approx0.22d_{\rm FA}$ for the \ac{cca} and $\approx0.19d_{\rm FA}$ for the \ac{csa}.

Hence, only a fraction of the radiative near-field ($\approx20\%$) can support finite-depth Bessel-beam operation, with the exact limit determined by the array geometry, aperture size, and carrier wavelength. The performed computer simulations cross-verify the analytical results and also confirm that the limits for the discrete half-wavelength spaced arrays closely follow the ones established analytically under the continuous array approximation. The results also show that, despite these limits, the usable non-diffracting range of a Bessel beam at mmWave and \ac{thz} frequencies still spans over up to several tens of meters. Our findings thus provide both theoretical limits and engineering insights on how to optimize the Bessel beam parameters for future near-field communication systems.

\appendices

\section{Derivation of $G^{c}_{\mathrm{norm}}(\mu_c,\infty)$~\eqref{eq:ginfty_real}}
\label{app:limit}
Starting from \eqref{eq:gnorm_closed_form}, the far-side behavior is obtained by expanding the two exponential terms and the Fresnel-integral differences for $\tau\to\infty$. This gives
\begin{equation}
\footnotesize
\begin{aligned}
G^{c}_{\mathrm{norm}}(\mu_c,\tau_c)
&=
(8\tau_c)^2
\left|
\frac{e^{-j\frac{\pi}{2}\mu^2\tau_c}}{\tau_c}
\frac{
\left(1-j\frac{\pi\mu_c}{2}\right)
e^{j\frac{\pi\mu_c}{2}}-1
}
{\pi^2\mu_c^2}
+
\mathcal{O}\left(\tau_c^{-2}\right)
\right|^2 
\end{aligned}
\label{eq:gnorm_far_side_expansion}
\end{equation}
Since $|e^{-j\frac{\pi}{2}\mu_c^2\tau_c}|=1$, the oscillatory phase does not affect the magnitude. Therefore, we get
\begin{equation}
\begin{aligned}
G^{c}_{\mathrm{norm}}(\mu_c,\infty)
&=
64
\left|
\frac{
\left(1-j\frac{\pi\mu_c}{2}\right)
e^{j\frac{\pi\mu_c}{2}}-1
}
{\pi^2\mu_c^2}
\right|^2 .
\end{aligned}
\label{eq:ginfty_mu_form}
\end{equation}
Defining $\xi=\pi\mu_c/2$, \eqref{eq:ginfty_mu_form} can be written as
\begin{equation}
G^{c}_{\mathrm{norm}}(\mu_c,\infty)
=
4
\left|
\frac{
e^{j\xi}(1-j\xi)-1
}
{\xi^2}
\right|^2 .
\label{eq:ginfty_complex_form}
\end{equation}
Thus,  we can obtain the real-form in~\eqref{eq:ginfty_real} using~\eqref{eq:ginfty_complex_form}.

\section{Derivation of $G^{s}_{\mathrm{norm}}(\mu_s,\infty)$~\eqref{eq:square_ginfty_complex_form_main}}
\label{app:square_limit}

Starting from \eqref{eq:square_gnorm_semiclosed_form}, the far-side behavior is obtained by expanding the two exponential terms and the Fresnel-integral differences for $\tau_s\to\infty$. For a fixed $\phi$, let $a_s(\phi)=\mu_s/2\cos(\phi).$
Using the large-$\tau_s$ expansion of the corresponding integrand, we obtain $G^{s}_{\mathrm{norm}}(\mu_s,\tau_s)$
\begin{equation}
\tiny
\begin{aligned}
&
(16\tau_s)^2
\Bigg|
\frac{
e^{-j\pi\mu_s^2\tau_s}
}{8\tau_s}
\int_0^{\pi/4}
\sec^2\phi
\left[
\frac{
e^{j\pi a_s(\phi)}
}{
j\pi a_s(\phi)
}
\frac{
1-e^{j\pi a_s(\phi)}
}{
\left(j\pi a_s(\phi)\right)^2
}
\right]
d\phi
+
\mathcal{O}\left(\tau_s^{-2}\right)
\Bigg|^2 .
\end{aligned}
\label{eq:square_gnorm_far_side_expansion}
\end{equation}
Since $|e^{-j\pi\mu_s^2\tau_s}|=1$, the oscillatory phase does not affect the magnitude. Therefore, we get
\begin{equation}
\small
\begin{aligned}
G^{s}_{\mathrm{norm}}(\mu_s,\infty)
&=
4
\left|
\int_0^{\pi/4}
\sec^2\phi
\left[
\frac{
e^{j\pi a_s(\phi)}
}{
j\pi a_s(\phi)
}
+
\frac{
1-e^{j\pi a_s(\phi)}
}{
\left(j\pi a_s(\phi)\right)^2
}
\right]
d\phi
\right|^2.
\end{aligned}
\label{eq:square_ginfty_mu_form}
\end{equation}
Let $ \xi_s(\phi)=\pi a_s(\phi)=\frac{\pi\mu_s}{2\cos(\phi)}$, then \eqref{eq:square_ginfty_mu_form} can be written as
\begin{equation}
\begin{aligned}
G^{s}_{\mathrm{norm}}(\mu_s,\infty)
&=
4
\left|
\int_0^{\pi/4}
\sec^2\phi
\left[
\frac{
e^{j\xi_s(\phi)}
}{
j\xi_s(\phi)
}
+
\frac{
1-e^{j\xi_s(\phi)}
}{
\left(j\xi_s(\phi)\right)^2
}
\right]
d\phi
\right|^2.
\end{aligned}
\label{eq:square_ginfty_complex_form}
\end{equation}
Equivalently, by combining the two terms inside the integral, we obtain
\begin{equation}
\begin{aligned}
G^{s}_{\mathrm{norm}}(\mu_s,\infty)
&=
4
\left|
\int_0^{\pi/4}
\sec^2\phi
\frac{
e^{j\xi_s(\phi)}
\left[
1-j\xi_s(\phi)
\right]
-1
}{
\xi_s^2(\phi)
}
d\phi
\right|^2.
\end{aligned}
\label{eq:square_ginfty_compact_form}
\end{equation}
Thus we can obtain~\eqref{eq:square_ginfty_complex_form_main}. Unlike the circular array case, the square array retains the angular integral because the array boundary depends on the azimuthal direction through the factor $\cos(\phi)$.
\section{Proof for Max gain being implicit}
\label{app:implicitgmax}
Define the inner term of
\eqref{eq:gnorm_closed_form} as $A(\tau_c)$ for a fixed $\mu_c$, so that $G^{c}_{\mathrm{norm}}(\mu_c,\tau_c)=64\tau_c^2 |A(\tau_c)|^2$. Then
\begin{equation}
    \frac{\partial G^{c}_{\mathrm{norm}}}{\partial \tau_c}
    =
    128\tau_c |A(\tau_c)|^2
    +
    128\tau_c^2
    \Re\left\{
    A^*(\tau_c)\frac{dA(\tau_c)}{d\tau_C}
    \right\}.
    \label{eq:gnorm_derivative}
\end{equation}
Therefore, a stationary point $\tau_c^\star$ satisfies
\begin{equation}
    |A(\tau_c^\star)|^2
    +
    \tau_c^\star
    \Re\left\{
    A^*(\tau_c^\star)
    \frac{dA(\tau_c)}{d\tau_c}
    \bigg|_{\tau_c=\tau_c^\star}
    \right\}
    =
    0.
    \label{eq:stationary_condition}
\end{equation}
Because $A(\tau_c)$ contains Fresnel-integral differences with $\tau_c$-dependent
arguments, \eqref{eq:stationary_condition} does not yield a clean closed-form
expression for $\tau_c^\star$ or $G_{\max}(\mu_c)$ for the \ac{cca} case. A similar argument applies to~\eqref{eq:square_gnorm_semiclosed_form}; hence, a closed-form solution for $G_{\max}(\mu_s)$ is also unavailable for the \ac{csa}.

\section{Numerical Accuracy of the \ac{csa} Angular Integral}
\label{app:csa_nphi_accuracy}
The \ac{csa} normalized array-gain expression~\eqref{eq:square_gnorm_semiclosed_form} contains a one-dimensional angular integral, which is evaluated numerically using \(N_{\phi}\) samples. This appendix validates the choice \(N_{\phi}=8001\) adopted throughout the paper by comparing the resulting Condition~A1 critical angle, \(\theta^{s}_{\rm crit,A1}\), with both a high-resolution numerical reference computed using \(N_{\phi}=128001\) and the analytical approximation reported in Prop~3a.

\begin{figure}[!t]
\centering
\includegraphics[width=0.96\columnwidth]{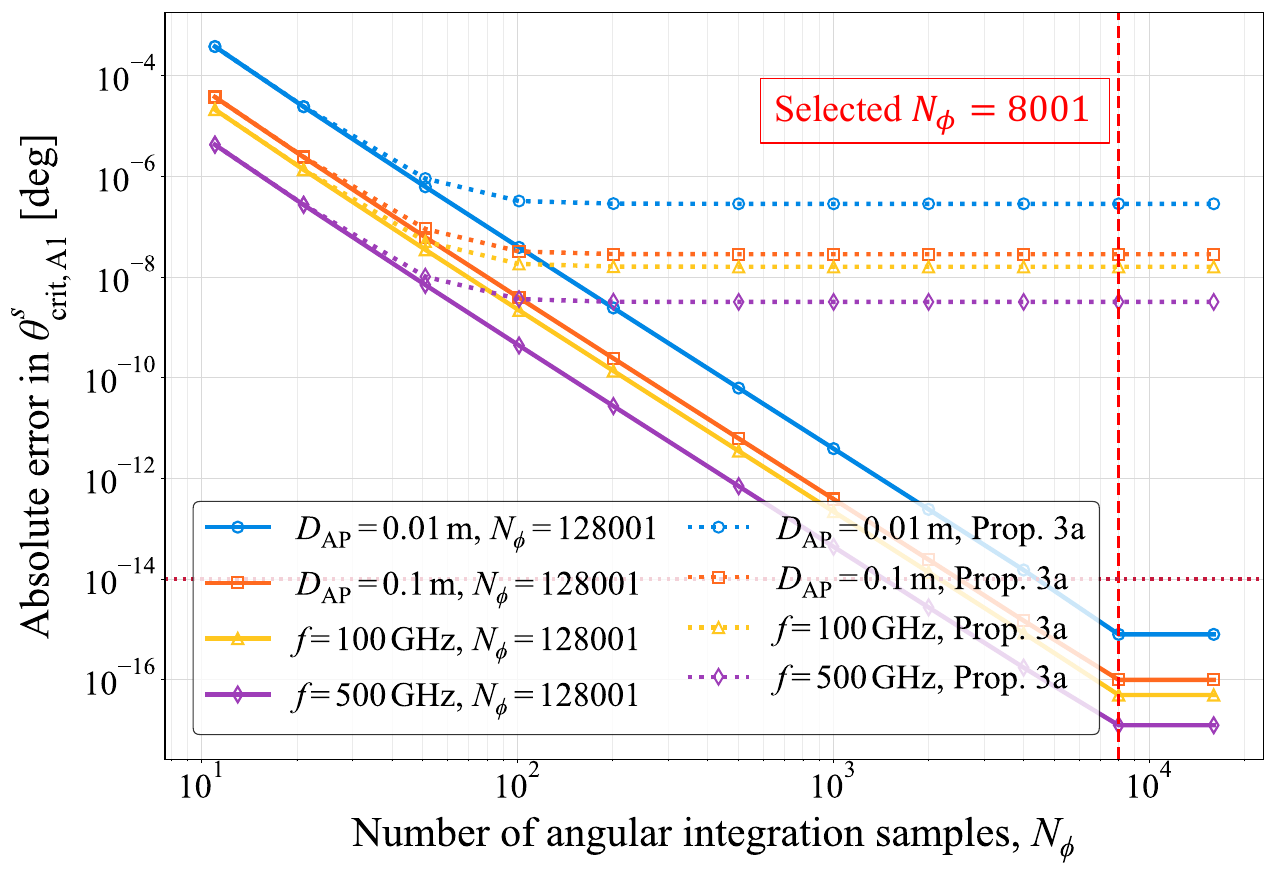}

\caption{Absolute error in \(\theta^{s}_{\rm crit,A1}\) versus the number of angular-integration samples \(N_{\phi}\), for \(D_{\rm AP} = \{0.01,0.1\}~\mathrm{m}\) with \(f=300~\mathrm{GHz}\), and \(f = \{100, 500\}~\mathrm{GHz}\) with \(D_{\rm AP}=0.3~\mathrm{m}\) (relative errors are $<10^{-5}\%$).}
\label{fig:error_3a_csa}
\end{figure}

Fig.~\ref{fig:error_3a_csa} illustrates the convergence of \(\theta^{s}_{\rm crit,A1}\) for different array sizes and operating frequencies. The solid curves show the error relative to the high-resolution numerical reference, while the dotted curves show the error relative to Prop.~3a. At \(N_{\phi}=8001\), the numerical integration error remains below \(10^{-14}\) degrees for all considered configurations, whereas the error relative to Prop.~3a is below \(10^{-6}\) degrees and is dominated by the rounding of the analytical constant rather than the numerical integration. Therefore, \(N_{\phi}=8001\) provides a practical tradeoff between accuracy and computation time when evaluating~\eqref{eq:square_gnorm_semiclosed_form} and the critical cone angles.

\section{Numerical Peak-Finding Algorithm}
\label{app:peak_finder}
\begin{algorithm}[t]
\caption{Peak-Finding Algorithm}
\label{alg:peak_finder}
\begin{algorithmic}[1]
\STATE \textbf{Input:} Array type $\kappa\in\{c,s\}$, array size, $\lambda$, $\theta$, $N_{\rm coarse}$, $\epsilon_z$, and $N_\phi$ for the \ac{csa}.
\STATE \textbf{Output:} $z_{\rm peak}$ and $G_{\rm peak}$.
\STATE Compute $d_{\rm FA}^{\kappa}$, $Z_{\max}^{\kappa}$, and define $G(z)=G_{\rm norm}(\theta,z)$ using the corresponding \ac{cca} or \ac{csa} expression.
\STATE Set $z_{\min}$ and $z_{\max}$.
\STATE Generate a logarithmic grid $\{z_n\}_{n=1}^{N_{\rm coarse}}$ over $[z_{\min},z_{\max}]$.
\STATE Evaluate $G_n=G(z_n)$ for all $n$.
\STATE Find $i^\star=\arg\max_n G_n$.
\IF{$i^\star=1$}
    \STATE $\mathcal B\leftarrow[z_1,z_2]$.
\ELSIF{$i^\star=N_{\rm coarse}$}
    \STATE $\mathcal B\leftarrow[z_{N_{\rm coarse}-1},z_{N_{\rm coarse}}]$.
\ELSE
    \STATE $\mathcal B\leftarrow[z_{i^\star-1},z_{i^\star+1}]$.
\ENDIF
\STATE Solve $z_{\rm peak}
    =
    \arg\min_{z\in\mathcal B}\left[-G(z)\right],$
using bounded scalar optimization with tolerance $\epsilon_z$.
\STATE Set $G_{\rm peak}=G(z_{\rm peak})$.
\STATE \textbf{return} $z_{\rm peak}$, $G_{\rm peak}$, $z_{\rm peak}/d_{\rm FA}^{\kappa}$, and $z_{\rm peak}/Z_{\max}^{\kappa}$.
\end{algorithmic}
\end{algorithm}
\begin{figure*}[t!]
\centering

\begin{minipage}[t]{0.49\textwidth}
\centering
\setcounter{subfigure}{0}

\subfloat[Varying $R$ and $f$, \(\theta=0.1^{\circ}\)]{%
    \includegraphics[width=0.485\linewidth]{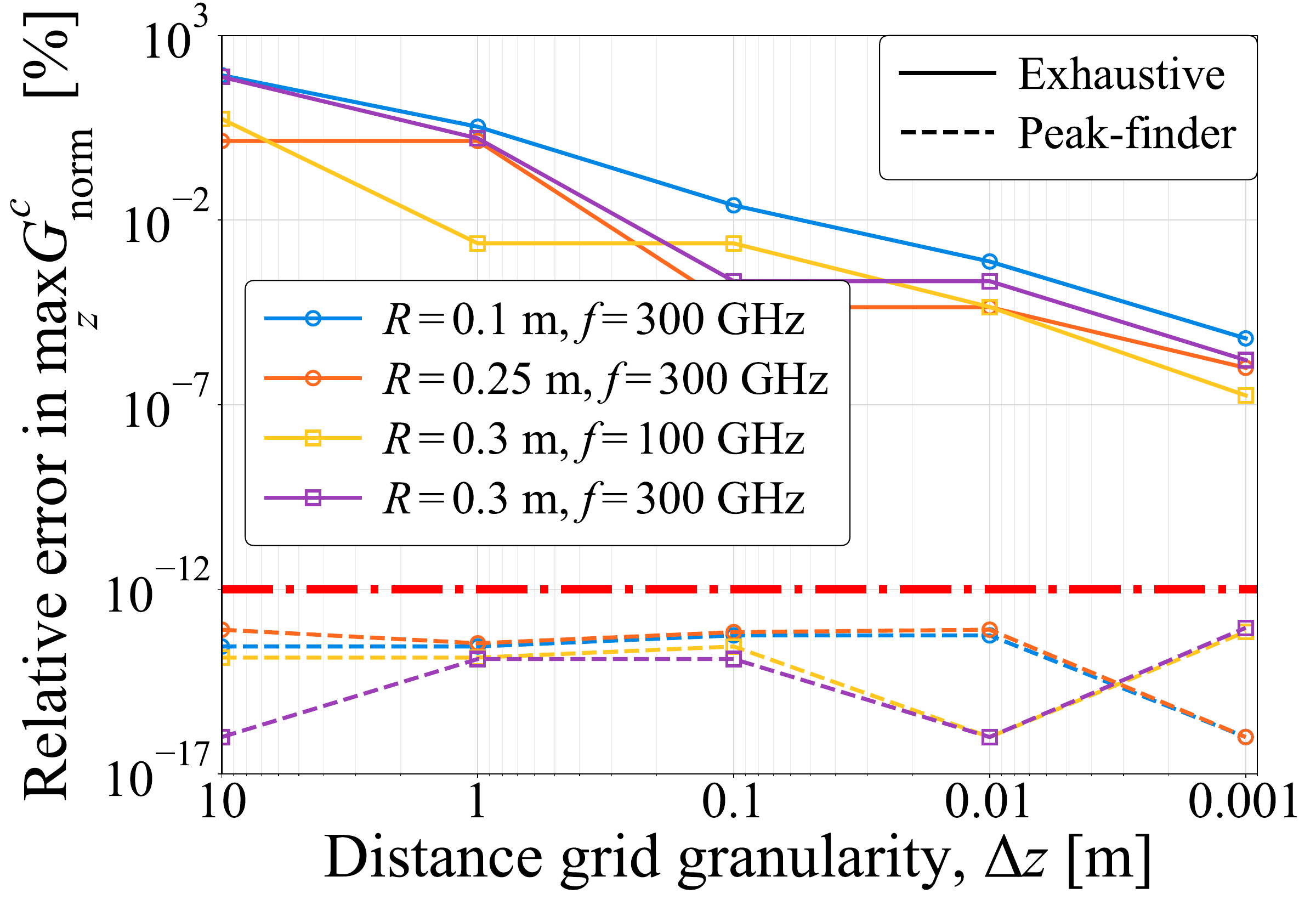}%
    \label{fig:peak_finder_CCA_R}}
\hfill
\subfloat[Varying \(\theta\), at \(300~\mathrm{GHz}\) and \(0.3~\mathrm{m}\)]{%
    \includegraphics[width=0.485\linewidth]{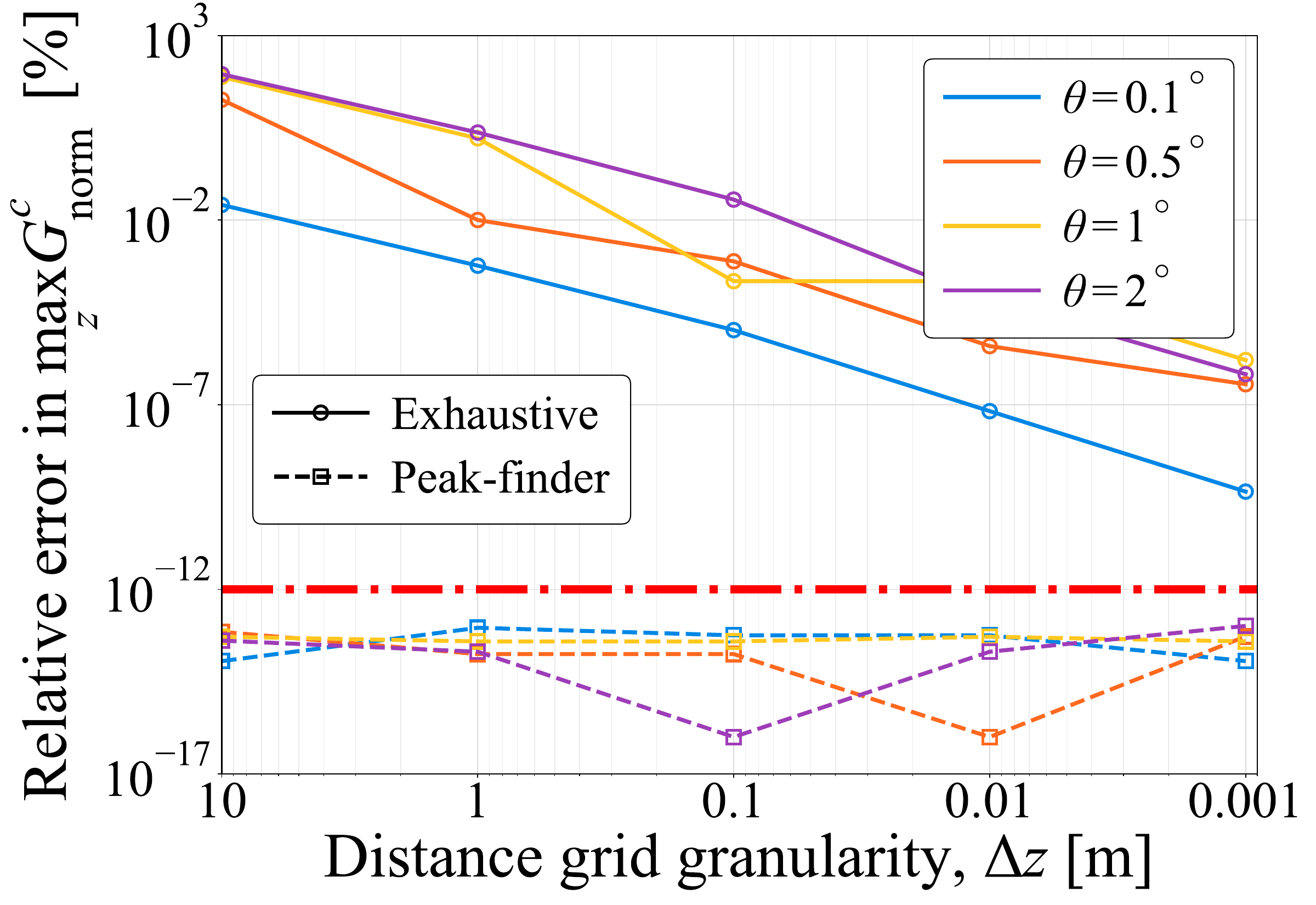}%
    \label{fig:peak_finder_CCA_f}
}

\begingroup
\makeatletter
\def\@captype{figure}
\caption{Peak-gain error versus distance-grid granularity for the \ac{cca}.}
\label{fig:peak_finder_CCA}
\makeatother
\endgroup

\end{minipage}
\hfill
\begin{minipage}[t]{0.49\textwidth}
\centering
\setcounter{subfigure}{0}

\subfloat[Varying $D_{\rm AP}$ and $f$, \(\theta=0.1^{\circ}\)]{%
    \includegraphics[width=0.485\linewidth]{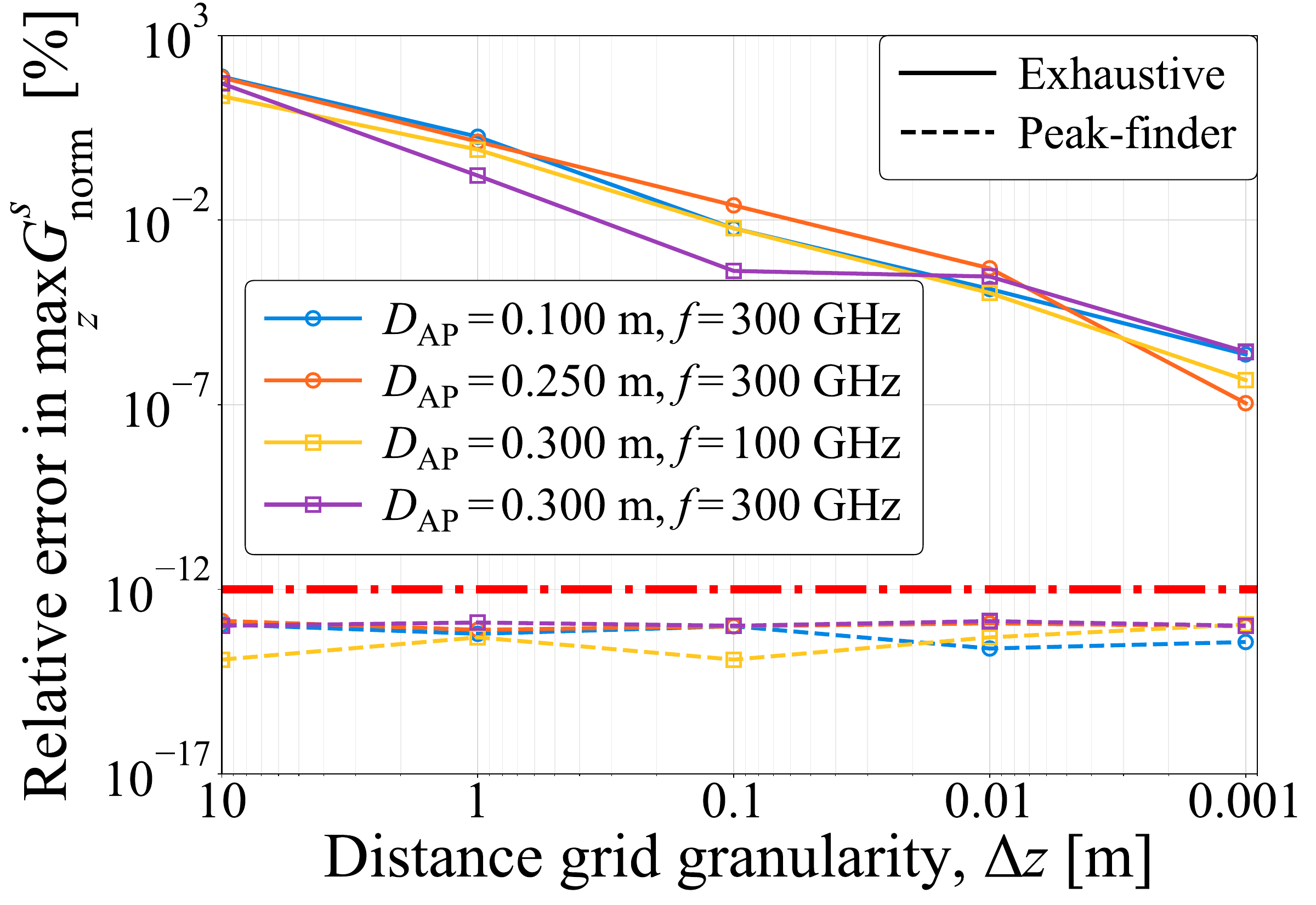}%
    \label{fig:peak_finder_CSA_DAP}
}
\hfill
\subfloat[Varying \(\theta\), at \(300~\mathrm{GHz}\) and \(0.3~\mathrm{m}\)]{%
    \includegraphics[width=0.485\linewidth]{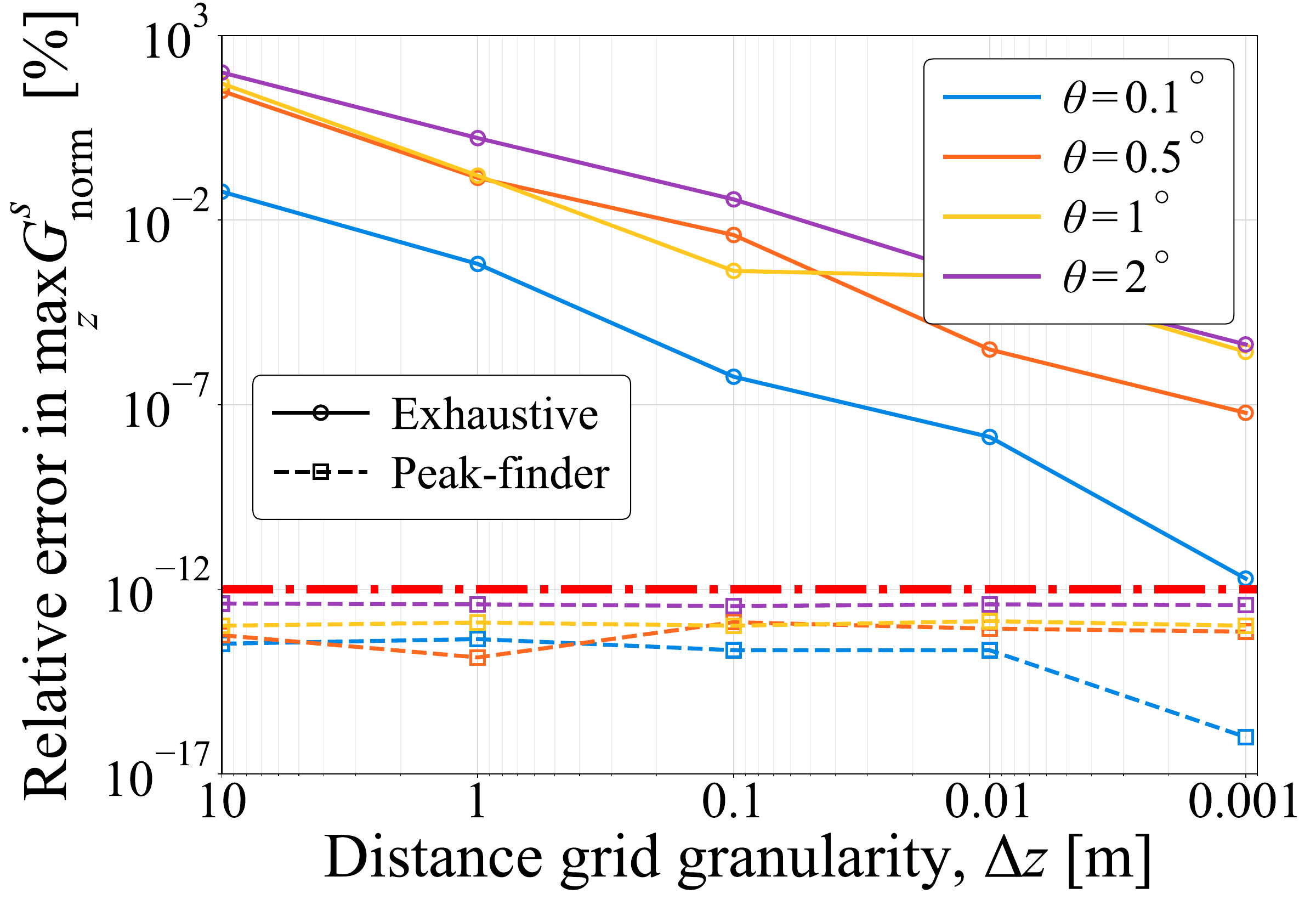}%
    \label{fig:peak_finder_CSA_f}
}

\begingroup
\makeatletter
\def\@captype{figure}
\caption{Peak-gain error versus distance-grid granularity for the \ac{csa}.}
\label{fig:peak_finder_CSA}
\makeatother
\endgroup

\end{minipage}

\vspace{-14pt}
\end{figure*}

Conditions~A2 and~B require the maximum normalized array gain over the axial \ac{ue} location for a given cone angle $\theta$, $G_{\rm peak}(\theta)=\max_{z\in[z_{\min},z_{\max}]}G_{\rm norm}(\theta,z).$ Since the corresponding stationary-point equation is implicit (see Appendix~\ref{app:implicitgmax}), the maximum is evaluated numerically. For the \ac{cca}, the closed-form expression~\eqref{eq:gnorm_closed_form} is used, while for the \ac{csa},~\eqref{eq:square_gnorm_semiclosed_form} is evaluated using $N_\phi=8001$ (see Appendix~\ref{app:csa_nphi_accuracy}). The search interval is selected as $z_{\min}=\max(d_{\rm RA}^{\kappa},10^{-8}Z_{\max}^{\kappa})$ and $z_{\max}=\max(3Z_{\max}^{\kappa},d_{\rm FA}^{\kappa})$, ensuring that the dominant Bessel peak is contained within the optimization region.

Algorithm~\ref{alg:peak_finder} employs a coarse-to-fine strategy. A coarse axial grid first identifies the dominant lobe, after which bounded optimization refines the peak location within the corresponding local bracket. Unlike exhaustive search, which refines the entire search interval, the proposed method performs high-resolution optimization only within the dominant lobe, substantially reducing the number of gain evaluations while eliminating the dominant grid-quantization error. Specifically, if $\Delta z_i$ denotes the local grid spacing and $\epsilon_z$ the optimizer tolerance, the local gain error satisfies $0
\leq
G_{\rm peak}
-
G_{\rm norm}(\theta,\hat z)
\leq
\frac{K_i}{2}\epsilon_z^2, $
where \(K_i\) denotes the maximum absolute curvature of \(G_{\rm norm}\) over the local bracket containing the dominant peak, and which is smaller than the exhaustive-search bound whenever $\epsilon_z<\Delta z_i/2$. Throughout this work, the optimization tolerance is set to $10^{-14}$ equivalent to relative error of $10^{-12} \%$.

The practical accuracy is evaluated in Fig.~\ref{fig:peak_finder_CCA} and~\ref{fig:peak_finder_CSA}. The reference solution is obtained by repeatedly refining candidate peaks until convergence. The solid curves correspond to direct exhaustive search, whereas the dashed curves show the proposed peak finder initialized with the same coarse grid. For both the \ac{cca} and \ac{csa}, the exhaustive-search error decreases with grid refinement, while the proposed method consistently remains below the prescribed numerical tolerance across all considered array sizes, operating frequencies, and cone angles. The slightly higher errors observed for the \ac{csa} originate from the numerical angular integration in~\eqref{eq:square_gnorm_semiclosed_form}. Overall, these results demonstrate that the proposed peak finder achieves substantially higher accuracy than direct exhaustive search at the same initial grid resolution.

\vspace{-1mm}
\balance
\bibliographystyle{IEEEtran}
\bibliography{ref}

\end{document}